\documentclass[aps,prd,preprintnumbers,twocolumn,amsmath,amssymb,longbibliography]{revtex4-1} 
\usepackage[utf8]{inputenc}
\usepackage[english]{babel}
\usepackage{graphicx}
\usepackage{subfiles}

\usepackage{url}
\usepackage{blindtext}
\usepackage{floatrow}
\usepackage{subcaption}
\usepackage{caption}
\usepackage{amssymb}
\usepackage{amsmath}
\usepackage{graphicx}
\usepackage{float}
\usepackage[braket, qm]{qcircuit}
\usepackage{tikz}
\usepackage{float}
\usepackage{appendix}
\usepackage{wrapfig}
\usepackage{xargs}
\usepackage{xcolor}
\usepackage[colorlinks]{hyperref}
\usepackage[capitalise]{cleveref}
\usepackage{import}
\usepackage{subfiles}

\definecolor{winered}{rgb}{0.8,0,0}
\definecolor{darkb}{rgb}{0,0,0.8}
\hypersetup{
    colorlinks=true,
    citecolor=blue,
    linkcolor=winered,
    filecolor=red,
    urlcolor=darkb,
}
\usepackage[margin=1in]{geometry}

\usepackage{subcaption}

\newcommand{\dket}[1]{|#1 \rangle}
\newcommand{\dbra}[1]{\langle #1 |}

\newcommand{\rcalrefs}[0]{{Bravyi2021_rcal,Nation_rcal,Nachman_rcal,Hicks_rcal,Wang_rcal,Geller_rcal,Maciejewski_rcal,van_den_Berg_rcal}}

\DeclareMathAlphabet{\pazocal}{OMS}{zplm}{m}{n}
\newcommand{\E}{\pazocal{E}}

\newcommand{\mc}[1]{\mathcal{#1}}
\newcommand{\mbb}[1]{\mathbb{#1}}

\begin{document}
\bibliographystyle{apsrev4-1}

\title{Applying NOX Error Mitigation Protocols to Calculate Real-time Quantum Field Theory Scattering Phase Shifts}

\author{Zachary Parks$^1$ $^\ast$}
\author{Arnaud Carignan-Dugas$^2$}
\author{Erik Gustafson$^3$}
\author{Yannick Meurice$^4$}
\author{Patrick Dreher$^5$ }

\affiliation{$^1$ Department of Computer Science, North Carolina State University, Raleigh, NC  27695, USA }
\affiliation{$^{\ast}$Corresponding Author: email: zpparks@ncsu.edu}
\affiliation{$^2$ Keysight Technologies Canada, Kanata, ON K2K 2W5, Canada }
\affiliation{$^3$ Fermi National Accelerator Laboratory, Batavia, Illinois 60510, USA}
\affiliation{$^4$ Department of Physics and Astronomy, The University of Iowa, Iowa City, IA 52242, USA }
\affiliation{$^5$ Department of Electrical and Computer Engineering, North Carolina State University, Raleigh, NC  27695, USA}

\begin{abstract}

Real-time scattering calculations on a Noisy Intermediate Scale Quantum (NISQ) quantum computer are disrupted by errors that accumulate throughout the circuits. To improve the accuracy of such physics simulations, one can supplement the application circuits with a recent error mitigation strategy known as Noiseless Output eXtrapolation (NOX). We tested these error mitigation protocols on a Transverse Field Ising model and improved upon previous phase shift calculations.  Our proof-of-concept 4-qubit application circuits were run on several IBM quantum computing hardware architectures. We introduce metrics that show between 22\% and 73\% error reduction  for circuit depths ranging from 13 to 37 hard cycles, confirming that the NOX technique applies to circuits with a broad range of failure rates. We also observed an approximate 28\% improvement in the accuracy of the time delay calculation of the scattering phase shift. These observations on different cloud-accessible devices confirm that NOX improves performance even when circuits are executed in substantially  time-separated batches. Finally, we provide a heuristic method to obtain systematic error bars on the mitigated results, compare them with empirical errors and discuss their effects on phase shift estimates.

\end{abstract}

\maketitle

\section{Introduction}

Recent advances in both quantum computing hardware platforms and software have raised expectations that in the coming years, quantum computers can be applied to problems that are not amenable to be solved using classical computers.  This has piqued the interest of physicists working in high energy, nuclear and condensed matter physics to consider applying this new computational technology toward physics problems that are inaccessible using classical computers.  Examples of such problems include modeling of high-energy particle physics real-time scattering experiments, dynamical processes in nuclear astrophysics describing neutron star evolution, and emergent low-energy phenomena in condensed matter systems \cite{https://doi.org/10.48550/arxiv.2204.08605,https://doi.org/10.48550/arxiv.2204.03381,https://doi.org/10.48550/arxiv.2203.07091,Klco:2021lap,Banuls:2019rao,Banuls:2019bmf,RevModPhys.86.153}.

Research focused on applying quantum computing toward these goals is already underway.  There have been several recent projects focused on real time evolution of the quantum Ising model (QIM) on a limited number of sites \cite{CerveraLierta2018exactisingmodel,Lamm:2018siq,GustafsonIsing,Gustafson:2019vsd, Kim_2020,yeteraydeniz2021scattering,vovrosh2020confinement,Salath__2015,Tan_2021,Smith_2019, Labuhn_2016,2017Natur.551..601Z,PhysRevE.58.5355,2017Natur.551..579B,2017Natur.551..601Z,Ciavarella:2022tvc}, as well as more complicated high energy and nuclear physics Hamiltonians. \cite{Alexandru_2019,Zohar_2013, Martinez_2016, Buyens:2016hhu,Jahin_2022,Klco:2018kyo,Brower2020lattice, karpov2020spatiotemporal,Holland_2020,Roggero_2020,surace2021scattering,Kharzeev:2020kgc,Ikeda:2020agk,Briceno:2020rar,PhysRevResearch.3.013078,Banuls:2019bmf,Zohar:2012ay,Zohar:2012xf,Zohar:2013zla,Zohar:2014qma,Zohar:2015hwa,Zohar:2016iic,Klco:2019evd,Ciavarella:2021nmj,Bender:2018rdp,Liu:2020eoa,Hackett:2018cel,Alexandru:2019nsa,Yamamoto:2020eqi,Haase:2020kaj,Armon:2021uqr,PhysRevD.99.114507,Bazavov:2015kka,Zhang:2018ufj,Unmuth-Yockey:2018ugm,Unmuth-Yockey:2018xak,Kreshchuk:2020dla,Kreshchuk:2020aiq,Raychowdhury:2018osk,Raychowdhury:2019iki,Davoudi:2020yln,Wiese:2014rla,Luo:2019vmi,Mathis:2020fuo,Singh:2019jog,Singh:2019uwd,Buser:2020uzs,Bhattacharya:2020gpm,Barata:2020jtq,Kreshchuk:2020kcz,Ji:2020kjk,Bauer:2021gek,Gustafson:2021qbt,Hartung:2022hoz,Grabowska:2022uos,Murairi:2022zdg,Gustafson:2022xlj,Farrell:2022vyh,Gustafson:2022xdt,McDonough:2022dtx, Chen_2022}.
For the QIM, phase shifts have been calculated from  the time delay of a wave packet due to interactions, using the real-time evolution in the early and intermediate stages of the collision~\cite{Gustafson_2021}. These calculations were performed with four qubits on superconducting and trapped ion quantum computing platforms. Recent work using simple error suppression methods \cite{Gustafson:2019vsd} and significantly larger Trotter steps \cite{GustafsonIsing,ybook,treview} than would be suggested by rigorous bounds, have provided reasonable extrapolations for scattering times up to the order of the approximate periodicity of the problem.

As with all computations run on today's quantum computing hardware platforms the problem of noise degrading the fidelity of the results must be addressed.  Various proposed noise mitigation approaches include probabilistic error cancellation methods \cite{temme2017shortdepth, ferracin2022, vandenBerg2023}, self-verifying circuits \cite{PhysRevD.106.074502, Farrell_2023, PhysRevLett.127.270502}, learning-based methods \cite{PRXQuantum.2.040330}, subspace expansion techniques \cite{McClean2020, PhysRevX.10.011004} and zero-noise extrapolation (ZNE). This paper focuses on a recent enhancement to zero-noise extrapolation methods called Noiseless Output eXtrapolation (NOX).

Zero-noise extrapolation (ZNE) is a widely used error mitigation technique due to its relatively simple and hardware-agnostic implementation \cite{PhysRevX.8.031027, Kandala_2019, Dumitrescu:2018njn, temme2017shortdepth, osti_1658010, li2017efficient, PhysRevA.105.042406, Kim2023, b4FTKim2023}. The main operating principle behind the ZNE protocol is to measure the observables of a target circuit and a family of equivalent circuits that have been noise-amplified in a controlled manner such that the measurements can be used to extrapolate to the zero-noise limit \cite{temme2017shortdepth}. 

A common way to amplify the noise in the ZNE protocol is by identity insertion, where a particular gate $U_{j}$ is replaced by $U_{j}(U^{\dagger}_{j}U_{j})^{a_{j}}$ for some integer $a_{j}$. The Fixed Identity Insertion method (FIIM) is the traditional insertion method where every gate $U$ is replaced by $r = 2a+1$ copies of itself to increase the total error in the circuits \cite{Dumitrescu:2018njn}. The Random Identity Insertion Method (RIIM) is a more gate-efficient insertion method, with a random integer $a$ chosen for each replaced gate such that fewer additional gates are needed for each amplified circuit \cite{He2020ZNERIIM}. 

The recently published NOX error mitigation strategy has been shown to enhance the performance of quantum circuits composed of noisy cycles of gates \cite{ferracin2022}.  
In NOX, errors are amplified via the local identity insertion of randomly compiled cycles. Randomized compilation effectively tailors errors to be purely stochastic, ensuring a linear error propagation regime. Randomized compiling is an important component of error mitigation because the extrapolation step often relies on the assumption of decoherent error propagation principles. Note that error mitigation could also apply for coherent error sources, but it would necessitate more circuits to account for the non-linear error propagation of coherent errors.  We have implemented this NOX procedure and applied it to our real-time phase shift calculations in order to verify that there are significant improvements in this application on actual quantum computing hardware architectures.

In \cref{sec:physics-bkgnd}, the physics background of the QIM is summarized, and the structure of the application circuit that will be analyzed is discussed.  \cref{sec:quantum-circuit} discusses the quantum circuit implemented for these computations.  \cref{sec:error-suppress} provides a detailed description of the NOX method.  \cref{sec:results} discusses the experimental results and provides a detailed error analysis based on the measured data.  Finally, \cref{sec:summary} summarizes the significance of the NOX technique and its general applicability for improving the error analysis for a broad range of algorithms run on quantum computing hardware platforms.

\section{Physics Background}
\label{sec:physics-bkgnd}

The QIM in one spatial dimension shown in~\cref{eq:HamiltonianIsingFree},
\begin{equation}
\label{eq:HamiltonianIsingFree}
\hat{H}_0 = - J \sum_{i=1}^{N - 1} \hat{\sigma}^x_i \hat{\sigma}^x_{i+1}  - h_T \sum_{i=1}^N \hat{\sigma}^z_i.
\end{equation}
is very well understood \cite{RevModPhys.55.775} and has been successfully implemented on NISQ devices  \cite{CerveraLierta2018exactisingmodel,Lamm:2018siq,GustafsonIsing,Gustafson:2019vsd}.  Because this model is constructed invariant under translations, the standard quantum mechanics two-particle-scattering problem can be reduced to a single-particle Schroedinger equation in an effective potential. In one spatial dimension, the simplest case of effective potential that can generate a phase shift for the reduced problem is a potential step adjacent to an infinite wall. This can be written as an interaction term:

\begin{equation}
\label{eq:Hamiltonianint}
\hat{H}_{int} = U \left(\frac{1-\hat{\sigma}^{z}_{N_{S}}}{2} \right)
\end{equation}

When constructing the initial wave packet, it is necessary to have some localization in space so that a distinct scattering event is visible. Because of this construct, the wave packet must have some momentum distribution because it is no longer a plane wave. We define the probabilities to be in the $\dket{\pm k}$ momentum state as
\begin{equation}
P_{\pm}(t)\equiv|\dbra{\pm k} \psi(t)\rangle |^2,
\end{equation}
and their normalized versions
\begin{equation}
\label{eq:rpminus}
R_{\pm}(t)\equiv \frac{P_{\pm}(t)}{P_ + (t)+P_-(t)}
\end{equation}
which by design satisfy
\begin{equation}
\label{eq:sumrule}
 R_+ + R_ -=1.
\end{equation}
The real-time evolution provides the time $t^\star$ necessary to reach the symmetric situation where $P_+(t^\star)=P_- (t^\star)$ and $R_-(t^\star)=0.5$. It is noted that the time $t^\star$ is determined by the symmetric condition $R_-(t^\star)=R_+(t^\star)=0.5$.  This also corresponds to the time where a classical particle would hit the wall. We can then compare $t^\star$ in the case where $U = 0$ and some non-zero value. We call these times $t^\star_{free}$
and $t^\star_{int.}$ respectively.
With this normalization from~\cref{eq:sumrule} the $R_+(t)$ and $R_-(t)$ get interchanged under time-reversal with respect to $t^\star$.  We define the difference
\begin{equation}
\label{eq:dstarb}
\Delta t^\star\equiv t^\star_{int.}-t^\star_{free},
\end{equation}
and argue that
\begin{equation}
\label{eq:dstar}
\Delta t^\star =\frac{\Delta t _W}{2}
\end{equation}
where $\Delta t_W$ is the Wigner time delay \cite{PhysRev.98.145}
\footnote{This relation can  be justified from the time-reversal argument that after $ t^\star_{int.}$ only {\it half} of the phase shift, $\delta(k)$, has built up while the other half builds after $t^\star$. This is why the total phase shift is historically denoted $2\delta(k)$.}.

Because of the small volume, we used a {\it deformed} sigmoid parametrization for $R_{-}(t)$ in~\cref{eq:R-minus}
\begin{equation}
R_{-}(t)\simeq A/(1+\exp(-(\frac{ t-\tilde{t}^\star}{w})))
\label{eq:R-minus}
\end{equation}
This parameterization will be used for the experimental results and analysis computations.

\section{The Quantum Circuit}
\label{sec:quantum-circuit}
The quantum circuit that implements the quantum field theory model of the scattering event in the system discussed previously can be described by three major components; state preparation, Trotterization, and the inverse quantum Fourier transformation (QFTr) measurement components. For these computations we focused on the errors generated by the two-qubit gates in this system.  

\begin{figure*}
\includegraphics[width=3.33in]{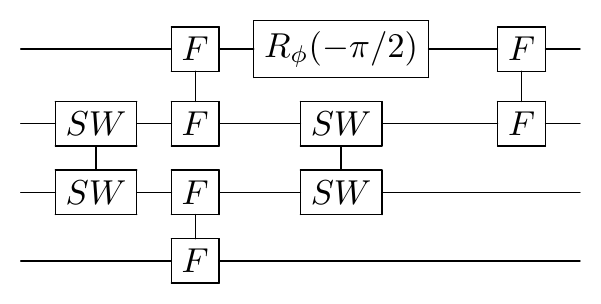}
\caption{Quantum circuit for preparing the measurement in the momentum space. We simplified the inverse quantum Fourier transform. This circuit maps the momentum states $\dket{+k}$ and $\dket{-k}$ into the $|0100 \rangle \langle 0100|$ and $|1000 \rangle \langle 1000|$ computational basis states, respectively. The entangling operations were transpiled using CNOTs as entanglers.}
\label{fig:qftrcirc}
\end{figure*}

\begin{figure*}[htpb]
    \begin{center}
    \includegraphics[width=.7\textwidth,
     clip=true,trim=0 0 0 0]{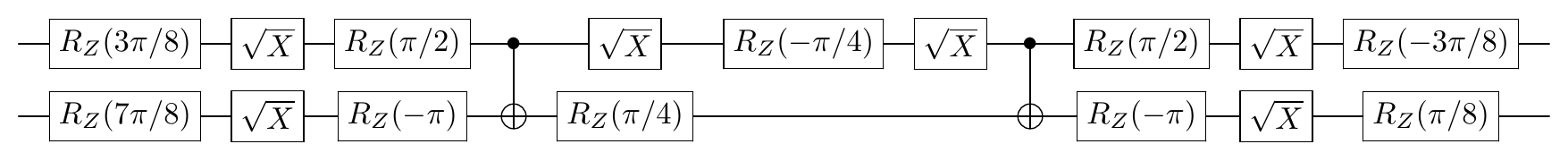}
     \end{center}
    \caption{Transpiled FF-gate}
    \label{fig:transpiled_F_gate}
\end{figure*}

\begin{figure*}
\includegraphics[width=3.33in]
{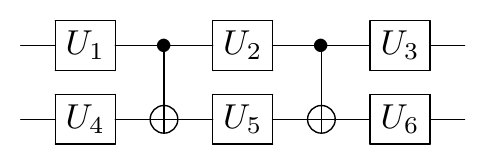}
\caption{Decomposition of FF-gate for IBM quantum computer.}
\label{fig:fgatedecomp}
\end{figure*}

As outlined in \cite{Gustafson_2021}, the state preparation circuit contains a single, two-qubit XX entangling operation. The Trotterization components consist of three XX entangling operations per Trotter time step.  The inverse QFTr measurement component of the application circuits consists of several SWAP and FF-gate entangling operations that perform the Fourier transformations (\cref{fig:qftrcirc}).  We note that the QFTr circuit is a simplified version as we were only interested in two momentum states and not all four momentum states in the system.  We provide a diagram of a transpiled F-gate circuit in \cref{fig:transpiled_F_gate} and a circuit diagram for the $F$-gate decomposition in terms of a set of defined U-gates in \cref{fig:fgatedecomp}.  The gates $U_i$ composing $F$ are defined using the Euler angle representation,
\begin{equation}
\label{eq:uigates}
    U_i \equiv U_i(\alpha_i, \beta_i, \gamma_i) = R_z(\alpha_i)R_x(\beta_i)R_z(\gamma_i).
\end{equation}
All of the U gates have an Euler angle rotation sequence of ZXZ except the U2 gate which is XZX.  The angles $\alpha_i$, $\beta_i$, $\gamma_i$ in the ZXZ rotation sequence for all U gates are listed in  \cref{tab:angles}. 

\begin{table}[!ht]
\caption{Angles for the rotation gates composing $U_i$ from Eq. (\ref{eq:uigates}) shown in Fig. \ref{fig:fgatedecomp}}.

\label{tab:angles}
\begin{tabular}{|c|c|c|c|}
\hline\hline
$i$ & $\alpha_i$ & $\beta_i$ & $\gamma_i$\\
\hline
1 & $-\frac{\pi}{4}$ & $\frac{\pi}{2}$ & $\pi$ \\
2 & $\pi$ & $\frac{3\pi}{4}$ & $-\frac{\pi}{2}$ \\
3 & $\frac{\pi}{2}$ & $\frac{\pi}{2}$ & $-\frac{3\pi}{4}$\\
4 & $-\frac{3\pi}{4}$ & $\frac{\pi}{2}$ & $\pi$\\
5 & $-\frac{3\pi}{4}$ & $\frac{\pi}{2}$ & $\pi$ \\
6 & $0$ & $0$ & $-\frac{\pi}{2}$\\
\hline\hline
\end{tabular}
\end{table}

It is noted that the number of two-qubit gates in these Fourier transforms expand quickly as a function of the number of qubits in the scattering problem.  As a result, the most challenging part of the computation was to mitigate the noise in those circuits to measure the systematic errors quantitatively. In addition, errors induced by imperfect Trotter steps were also challenging. These systematic errors are expected to dominate the error profile of the scattering phase shift compared to the statistical errors.  This project will be to utilize error mitigation methodologies that can quantitatively estimate these systematic errors.

\section{Error Suppression}
\label{sec:error-suppress}

We begin our discussion of error suppression with a brief review of the circuit model of a quantum computation and the formalism by which we investigate the error processes that affect such calculations. 

The circuit model of quantum computation focuses on the idea of a \emph{circuit} as an ordered list of instructions within a program.  Each instruction consists of a tuple containing the registers to address (e.g. the qubit indices) and the operation to perform. Possible operations include \emph{gates} (which are unitary operations on the register space), measurement, and state preparation (e.g. qubit reset). Finally, a computational \emph{cycle} (also known as a \emph{moment}) usually refers to parallel instructions, although it can also refer to a set of sequential instructions addressing disjoint sets of registers.

Error processes depend on the gates but can also affect additional registers beyond those that the instruction is directly meant to address. The errors also depend on the precise relative scheduling of the instructions during a cycle. Fortunately, it is generally observed that error channels only weakly depend on prior cycles. For this reason, we attach fixed error channels to cycles rather than to gates. That is, we express the error-prone version of the ideal cycle $C_i$ as $\mc E_i C_i$. The error channel $\mc E_i$ depends on the cycle $C_i$.

Quantum computers output results as probability distributions. The overarching idea behind error suppression (to contrast with error correction) is to pair noisy application circuits with other circuits and combine the probabilistic results in such a way that the desired observable estimate is closer to the ideal expectation value.

In this work, we implement three compounding error suppression techniques: readout calibration (RCAL) \cite{\rcalrefs}, randomized compiling (RC) \cite{2016efficienttwirling}, and Noiseless Output eXtrapolation (NOX) \cite{ferracin2022}.  The circuits' generation and the analysis related to those three mitigation techniques were performed by the True-Q software \cite{trueq}.  

\subsection{Readout Calibration and Randomized Compiling}
\label{sec:RCAL-RC-text}

RCAL is implemented using a few quantum circuits to estimate the noise matrix associated with the measurement step \cite{\rcalrefs}. The inverse noise matrix is then applied to subsequent outcome distributions, effectively suppressing measurement errors.  

The role of RC is to regulate the propagation of cycle errors throughout the computation. Often, the regulation effect substantially reduces the overall circuit failure rate over ungoverned circuits \cite{Hashim_2021,ville2021, Gu2022}.  The principle behind RC is to replace an application circuit that is meant to be run $N_{\rm shots}$ times with $n_{\rm rand}$ equivalent circuits sampled over a particular random distribution, each to be performed $N_{\rm shots}/n_{\rm rand}$ times (in our case we chose $N_{\rm shots}=10,000$ and $n_{\rm rand}=30$ circuit randomizations). A well-known effect of RC is that it effectively tailors general Markovian error sources into stochastic error channels. That is if we divide a circuit into appropriate cycles \footnote{In RC, the circuit subdivisions are called \emph{dressed cycles}, and must take a specific form. See \cite{2016efficienttwirling}.},
$\mc C_{\rm ideal}  = C_m\cdots C_2C_1$, the average RC circuit can be well approximated as a sequence of ideal cycles interleaved with Pauli stochastic error channels $\mc E_i$.
\begin{align}
 \left\langle \mc C \right\rangle_{\rm RC}  \simeq \mc E_mC_m\cdots \mc E_2C_2 \mc E_1 C_1 ,
\label{eq:Cavg-RC}
\end{align}
The action of $\mc E_i$ on a state $\rho$ is defined as
\begin{align}
    \mc E_i [\rho] = \sum_j {p_j(C_i)} P_j \rho P_j^\dagger~,
\label{eq:Pauli-Sto-Channel}
\end{align}
where $P_j$ are Pauli operators and $\{p_j(C_i)\}_j$ is an error probability distribution. {The error profile  $\{p_j(C_i)\}_j$ is proper to the noisy device under consideration and generally depends on the cycle $C_i$ since some operations are more error-prone than others.}

\subsection{NOX Error Mitigation Protocol}
\label{sec:NOX-text}

The RC protocol is the starting point for the NOX protocol implementation as described in \cite{ferracin2022}.   The protocol is implemented on the intended scattering circuit using the basic RC and enhanced jointly with a family of noise-amplified versions, each of which is randomized as follows via RC:
\begin{align}
\left\langle \mc C \right\rangle_{\rm RC} \simeq & ~\mc E_mC_m\cdots \mc E_2 C_2 \mc E_1 C_1~,  \tag{Non-amplified RC circuit}\\
\left\langle \mc C_1 \right\rangle_{\rm RC} \simeq & ~\mc E_mC_m\cdots \mc E_2C_2 \mc E_1^{1+\alpha} C_1 ~, \tag{$\mc E_1$-amplified circuit}\\
\left\langle \mc C_2 \right\rangle_{\rm RC} \simeq & ~\mc E_mC_m\cdots \mc E_2^{1+\alpha}C_2 \mc E_1 C_1 ~, \tag{$\mc E_2$-amplified circuit}\\
&\vdots \notag \\
\left\langle \mc C_m \right\rangle_{\rm RC} \simeq & ~\mc E_m^{1+\alpha}C_m\cdots \mc E_2C_2 \mc E_1 C_1~. \tag{$\mc E_m$-amplified circuit}
\end{align}
As indicated via the $\left\langle \cdot \right\rangle_{\rm RC}$ notation, each of the $m+1$ above circuits is averaged via RC to ensure that the effective error model remains Pauli stochastic.  In our case, the error amplification $\mc E_i \rightarrow \mc E_i^{1+\alpha}$ (denoted by the error amplification factor $\alpha$) is approximately obtained by replacing a cycle of parallel CX gates with a circuit-equivalent odd sequence of identical CX cycles, interleaved with randomly compiled Pauli operations. We
chose a repetition number of $11$ CX cycles, corresponding to an amplification parameter of $\alpha = 10$.

\begin{figure*}[t]
    \centering
    \includegraphics[width=0.7\textwidth]{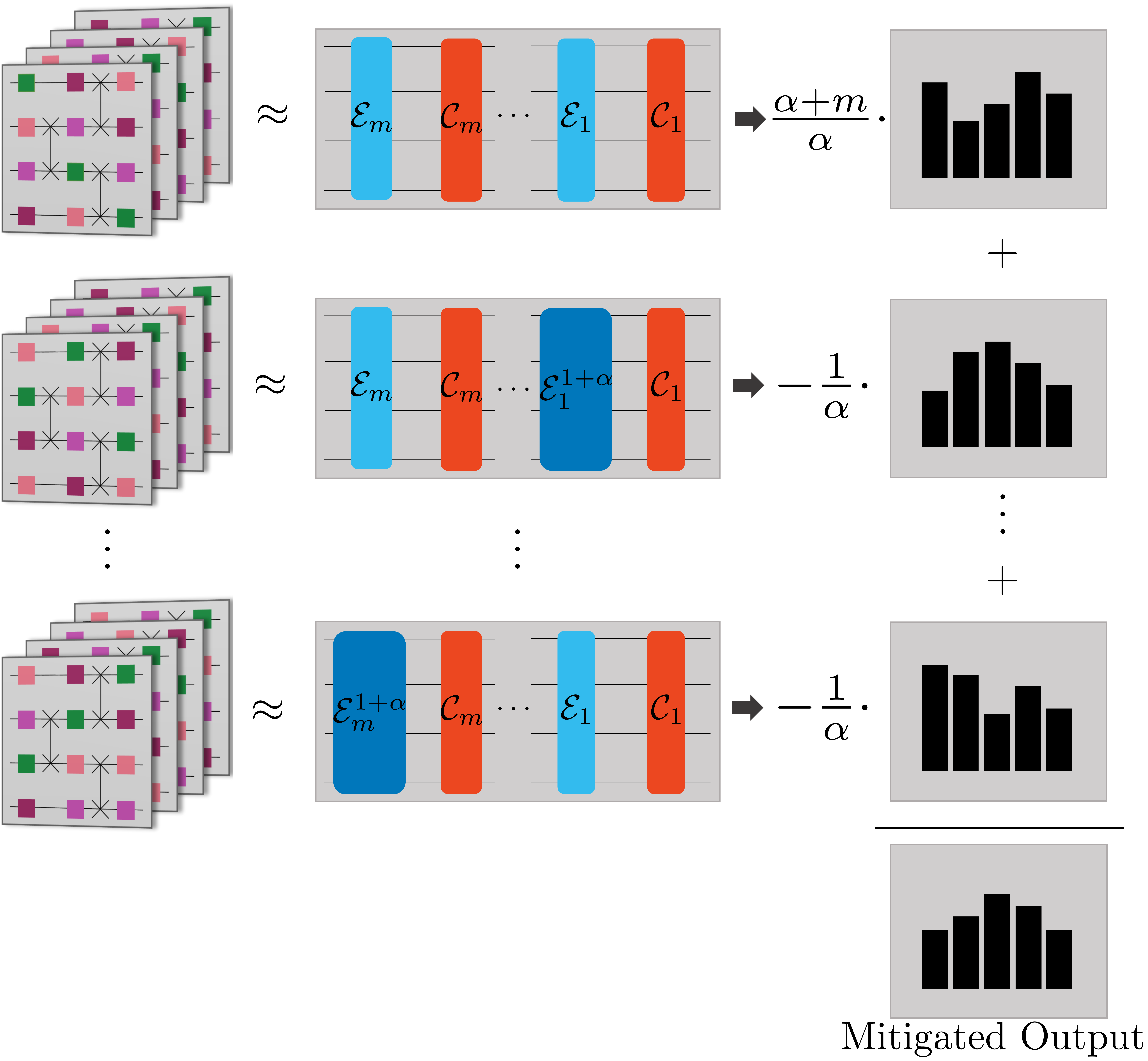}
     \caption{Overview of the NOX+RC procedure. The application circuit is run jointly with a family of noise-amplified versions of itself, each of which is averaged via RC (left). Each set of RC circuits can be approximated (middle) as a sequence of ideal cycles $\mc C_{i}$ (red) interleaved with non-amplified $\mc E_{i}$ (cyan) and amplified Pauli stochastic error channels $\mc E_{i}^{1+\alpha}$  (dark blue). The individual distributions of the original and of each error-amplified RC circuit (right) are used to calculate a mitigated expectation value of a given observable $O$ according to \cref{eq:nox}.
    }
   \label{fig:NOX}
\end{figure*}

Given an observable $O$, we denote its expected value given an effective RC circuit $\left\langle \mc C \right\rangle_{\rm RC}$ as $\mbb E (O | \left\langle \mc C \right\rangle_{\rm RC})$. With this notation at hand, the RC and NOX+RC estimates for $O$ are, respectively:
\begin{align}
    \hat{O}_{\rm RC} := & \mbb E (O | \left\langle \mc C \right\rangle_{\rm RC})~, \\
    \hat{O}_{\rm NOX+  RC}  := &\frac{\alpha+m}{\alpha} \mbb E (O | \left\langle \mc C \right\rangle_{\rm RC})\notag \\ &- \frac{1}{\alpha} \sum_{i=1}^m \mbb E (O | \left\langle \mc C_i \right\rangle_{\rm RC})~.
    \label{eq:nox}
\end{align}
This NOX+RC procedure is pictorially illustrated in~\cref{fig:NOX}. It is noted that this mitigation technique performs a first-order correction and is not expected to return mitigated results that perfectly match the ideal distribution even in the limit of infinite sampling.  

\subsection{First-order correction through NOX}
\label{sec:NON-1st-correct}

In this section, we provide a brief analysis of NOX, and show the correction that it provides on the unmitigated circuit.  First, let's re-express the effective
RC circuit as
\begin{align}\label{eq:rc_delta_appA}
    \langle \mc C\rangle_{\rm RC}  = &\mc C_{\rm ideal}  + \sum_{i=1}^m \Delta_i + \sum_{i>j}^m\sum_{j=1}^m \Delta_i \Delta_j \notag \\
    &+ \mathcal{O}\Bigg({m\choose3} \Delta^{3}\Bigg)
\end{align}
where
\begin{align}
    \Delta_i := C_m \cdots C_{i+1} (\mc E_i-1) C_i \cdots C_1~.
\end{align}
If the assumptions underlying NOX are perfectly respected, then we obtain amplified circuits of the form
\begin{align}\label{eq:nox_delta_horder}
    \langle \mc C_i\rangle_{\rm RC}  = \mc C_{\rm ideal}  &+ \alpha \Delta_i +\sum_{j=1}^m \Delta_j + \alpha \sum_{j > i}^m \Delta_j \Delta_i \notag \\
    &  +\alpha \sum_{i>j }^m \Delta_i \Delta_j+\sum_{k>j}^m\sum_{j=1}^m \Delta_k \Delta_j \\
    & + \mathcal{O}(m\Delta^{2}+m^{3}\Delta^{3})  ~,
\end{align}
where the higher-order terms include a $O(m^3 \Delta^3)$ scaling, but also a $O(  m \Delta^2)$ scaling from the Taylor expansion of $\mc{E}_i^{1+\alpha}$. If we substitute \cref{eq:rc_delta_appA,eq:nox_delta_horder} in  \cref{eq:nox}, we immediately get~\cref{eq:nox-result}

\begin{align}
  \left\langle  \mc C \right\rangle_{\rm NOX} &:= \frac{\alpha+m}{\alpha}\left\langle \mc C \right\rangle_{\rm RC} - \frac{1}{\alpha} \sum_{i=1}^m  \left\langle \mc C_i \right\rangle_{\rm RC} \notag \\
    &= \mc C_{\rm ideal} - \sum_{i>j}^m\sum_{j=1}^m \Delta_i \Delta_j+\mathcal{O}(m^{2}\Delta^{2})
\label{eq:nox-result}
\end{align}
which indicates a remaining bias that scales as $O(m^2\Delta^2)$, as opposed to $O(m \Delta)$ in the unmitigated case.

\section{Experimental Results and Analysis}
\label{sec:results}

The implementation of the NOX protocol was run on the superconducting IBM quantum hardware platforms  \texttt{ibmq\_kolkata}, \texttt{ibmq\_guadalupe} and \texttt{ibmq\_montreal}.  This section describes the experimental results and analysis using \texttt{ibmq\_kolkata} data.  The same analysis was also conducted with both the \texttt{ibmq\_guadalupe} and \texttt{ibmq\_montreal} data, The full set of results for all three platforms are summarized in \cref{tab:metrics}.

\subsection{The $\Delta R_{-}(t)$ calculations}
\label{sec:Delta R}

The project targeted qubits 19, 20, 22, and 25 on the 27-qubit \texttt{ibmq\_kolkata} quantum hardware platform. In post-processing, we examined the effects of each error mitigation procedure mentioned in \cref{sec:error-suppress} on the real-time evolution and calculation of the scattering phase shift. 

\begin{figure}[ht]
     \centering
     \begin{subfigure}{8cm}
     \includegraphics[width=8cm]{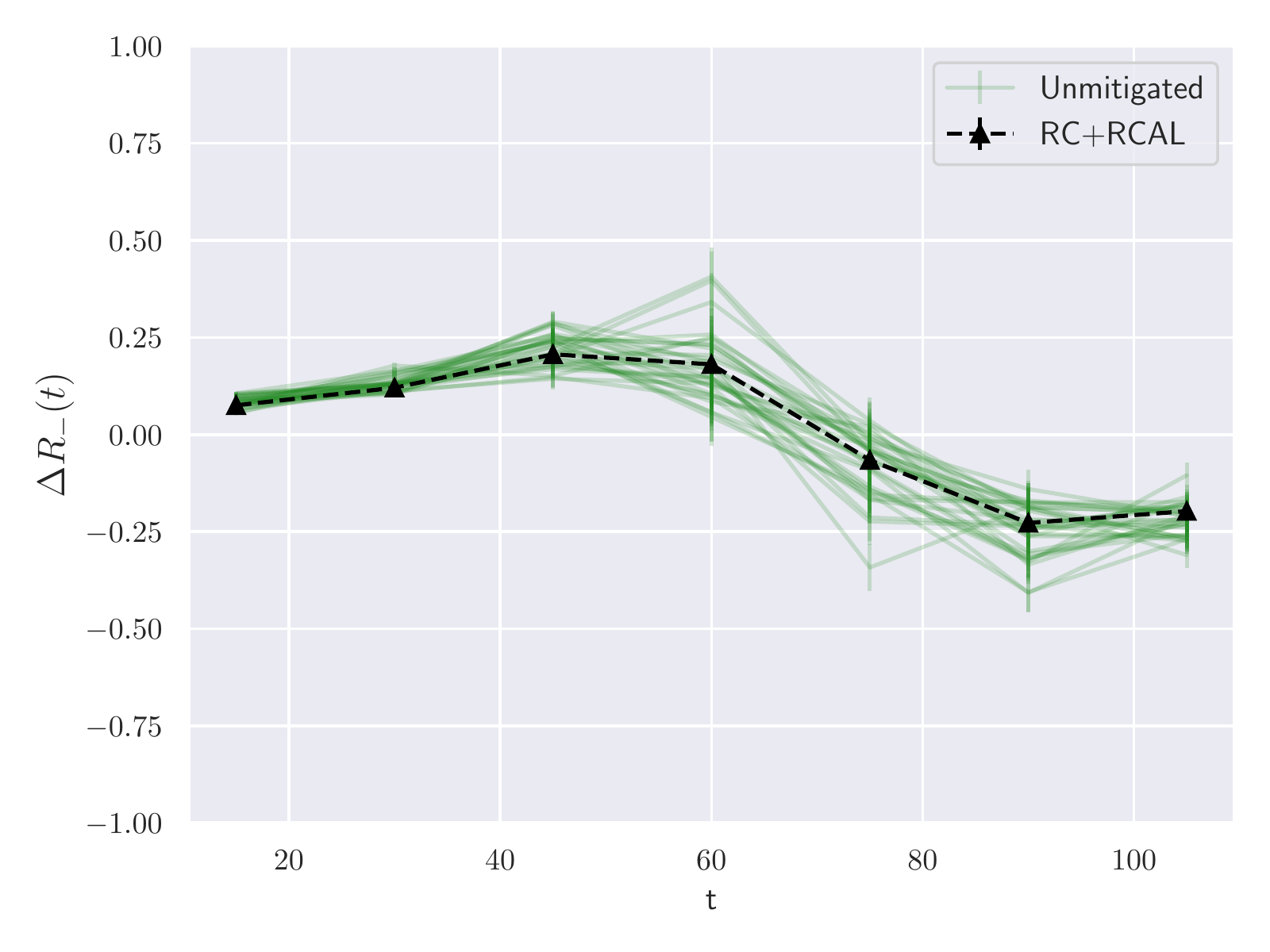}
     \caption{ free case }
     \label{fig:unmit_free}
     \end{subfigure}
     \begin{subfigure}{8cm}
     \includegraphics[width=8cm]{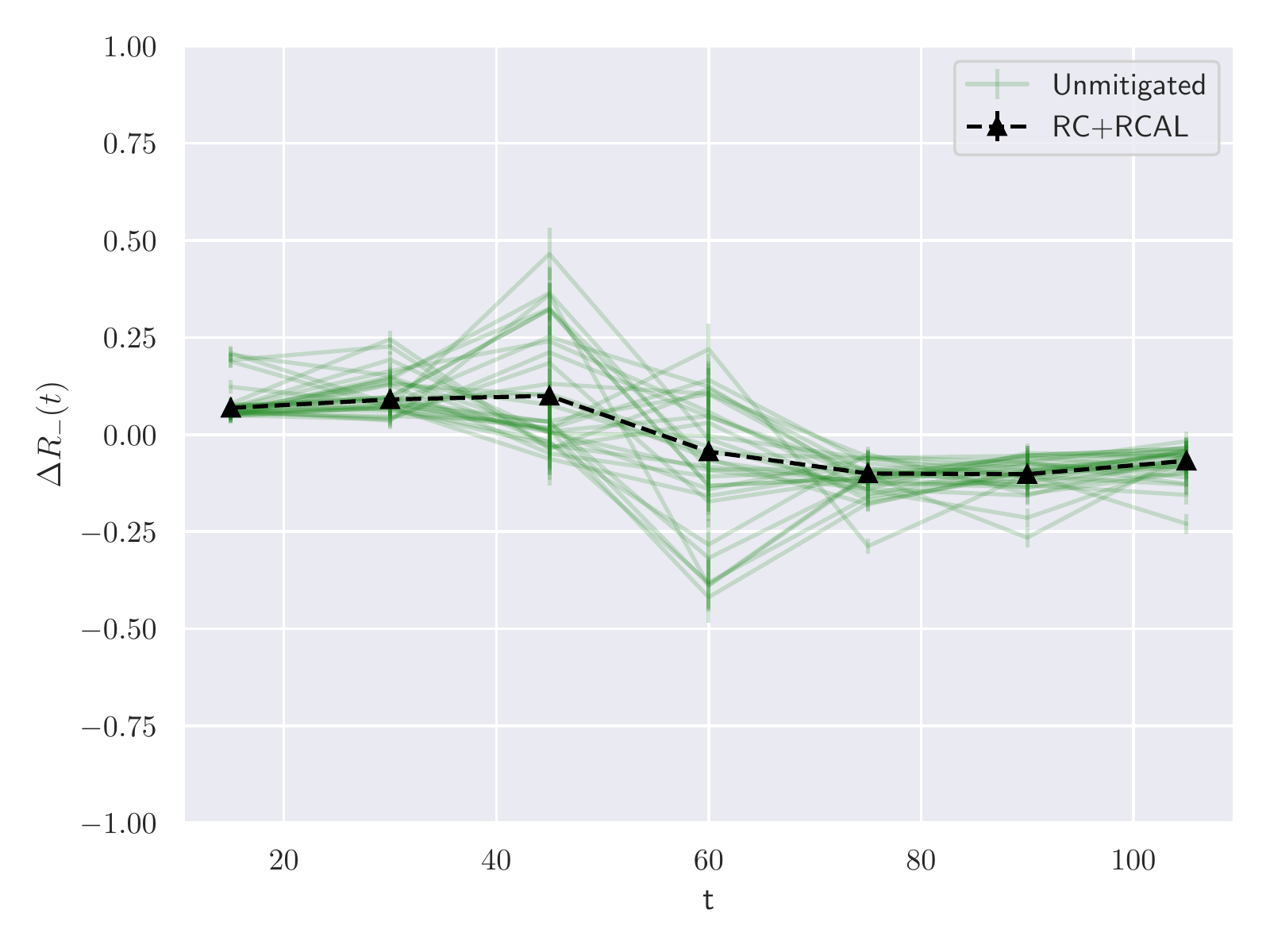}
    \caption{ interacting case }
     \label{fig:unmit_int}
     \end{subfigure}
     \caption{Difference in measured $R_{-}(t)$ from the ideal results versus the time evolution for the \texttt{ibmq\_kolkata} unmitigated data and with RC+RCAL applied to both the free and interacting cases}
    \label{fig:unmit_cloud_diff}
 \end{figure}

The first analysis of the \texttt{ibmq\_kolkata} data using combinations of these error mitigation procedures is shown in \cref{fig:unmit_cloud_diff}. 
This figure plots $\Delta R_{-}(t)$, defined as the deviation of the measured $R_{-}(t)$ from the ideal normalized reflection probability at evolution time $t$, for both the unmitigated data and the data obtained after applying the RC+RCAL error mitigation protocols.
 
We computed the unmitigated data for the free and interacting cases by individually selecting the 30 RC equivalent circuits used to calculate $R_{-}(t)$ at each Trotter time step. 
We then computed and plotted the 30 $\Delta R_{-}(t)$ data points at each time step and connected them to form 30 curves.
Next, we computed and plotted $\Delta R_{-}(t)$ obtained after applying the RC+RCAL protocols for the free and interacting cases. 
The results for $\Delta R_{-}(t)$ versus $t$ are shown for both the free \cref{fig:unmit_free} and interacting cases \cref{fig:unmit_int}. Comparing $\Delta R_{-}(t)$ obtained with and without mitigation protocols demonstrates the effectiveness of the combined RC+RCAL mitigation over unmitigated data.

The next analysis of the \texttt{ibmq\_kolkata} data applies the NOX procedure described in \cref{sec:NOX-text} and \cref{sec:NON-1st-correct}. 
As discussed in \cref{sec:NOX-text}, applying NOX requires adding $m+1$ additional circuits to the computation, corresponding to the number of hard cycles in the application circuit. 

Each of the circuit components described in \cref{sec:physics-bkgnd} contributes to the total number of hard cycles. There is one hard cycle in the state preparation component of the circuit that remains constant for every Trotter step of the simulation. The QFTr circuit component, by itself, consists of twelve hard gate cycles.  However, the transpilation optimization reduces this hard cycles count down to eight and also remains a constant factor for each Trotter step.  A single Trotter time step in our application requires six two-qubit entangling gates \cite{Gustafson_2021}.  During transpilation, this two-qubit gate count is reduced from six to four.  Therefore, Trotter step one requires a total hard cycle count of 13. Each subsequent Trotter step adds an 4 additional hard cycles.  We implemented a total of seven Trotter steps.  Therefore, the total number $m$ of hard gate cycles in our circuits as we perform the Trotterization from steps one to seven ranged from 13 to 37, covering a wide scope of circuit failure rates. 
 
In addition to these $m$ circuits, we include the original non-amplified circuit to analyze a total of $m+1$ circuits for each Trotter time step. The total number of circuits was sufficiently large that our experiments had to be batched in multiple jobs and effectively run on the cloud over many hours.  

We analyzed the NOX procedure by computing $\Delta R_{-}(t)$ obtained using NOX without applying RC and comparing it to values obtained using RC+RCAL and NOX+RC+RCAL.  This set of computations illustrates the significant improvement by combining NOX with the original RC+RCAL error mitigation.  As expected, the unmitigated results are amplified by NOX, and the values of $\Delta R_{-}(t)$ are larger than the unmitigated values alone. This is because the NOX procedure amplifies the unmitigated values shown in \cref{fig:unmit_cloud_diff} without the effects of RC present.

\begin{figure}[ht]
     \centering
     \begin{subfigure}{8cm}
     \includegraphics[width=8cm]{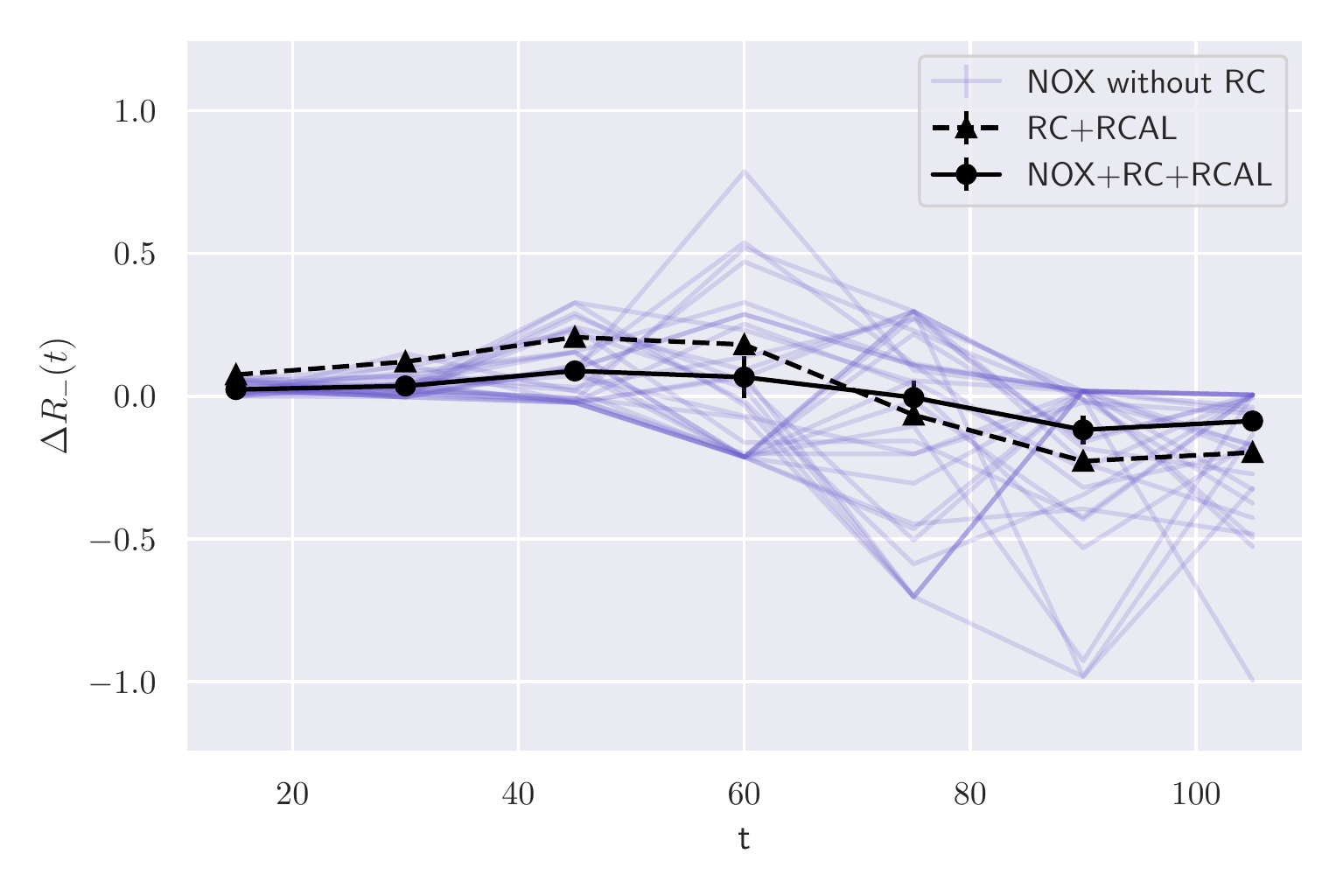}
     \caption{ free case }
     \label{fig:NOX_free}
     \end{subfigure}
     \begin{subfigure}{8cm}
     \includegraphics[width=8cm]{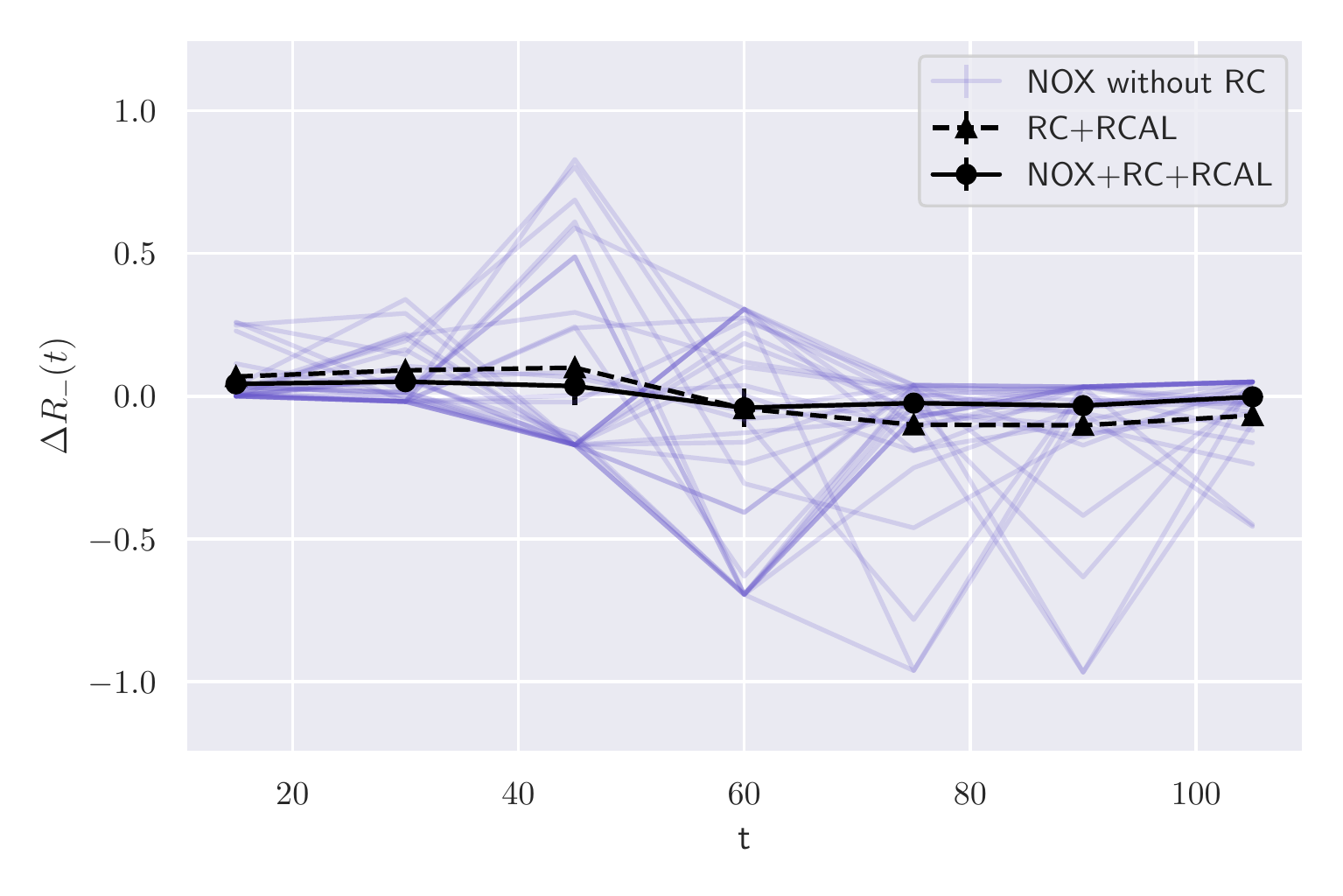}
    \caption{ interacting case }
     \label{fig:NOX_int}
     \end{subfigure}
     \caption{Difference in $R_{-}(t)$  from the ideal results versus the time evolution for NOX without RC, RC+RCAL, and NOX+RC+RCAL using the \texttt{ibmq\_kolkata} data}
    \label{fig:NOX_cloud_diff}
 \end{figure}

Normally, one would perform RC on those $m+1$ circuits and combine the $m+1$ RC averaged circuits using \cref{eq:nox} to compute the data point for each Trotter time-step. However, to compare NOX without RC, we batched the $30(m+1)$ circuits into 30 sets of circuits and then used \cref{eq:nox} to get 30 expectation values for $R_{-}(t)$ for each Trotter step. We then compute the 30 $\Delta R_{-}(t)$ data points for each Trotter step to generate the 30 curves labeled NOX without RC and plotted them in \cref{fig:NOX_cloud_diff}.

We then compute $\Delta R_{-}(t)$ for the entire time evolution after applying the NOX+RC+RCAL protocols and include these curves in \cref{fig:NOX_cloud_diff}. In addition, the RC+RCAL values from \cref{fig:unmit_cloud_diff} are re-plotted over the entire time evolution. This set of data is plotted for both the free (\cref{fig:NOX_free}) and the interacting cases (\cref{fig:NOX_int}). These graphs show that although the NOX procedure initially amplifies the unmitigated results and appears to make the error mitigation worse, the addition of NOX to RC+RCAL delivers an overall improved error mitigated result for $\Delta R_{-}(t)$ over the entire time evolution as compared to only applying RC+RCAL.  

These results clearly illustrate that the NOX method implemented alone without any additional mitigation did not show dramatic improvements in the $R_{-}(t)$ measurements. However, when the NOX method was combined with RC+RCAL, the final result improved upon just the RC+RCAL. This NOX+RC+RCAL combined error suppression provided the best error mitigation for having the measured data most closely follow the ideal evolution of $R_{-}(t)$ versus t and provided the highest accuracy among the methods tested when calculating $R_{-}(t)$.  

\begin{figure}[ht]
     \centering
     \begin{subfigure}{8cm}
     \includegraphics[width=8cm]{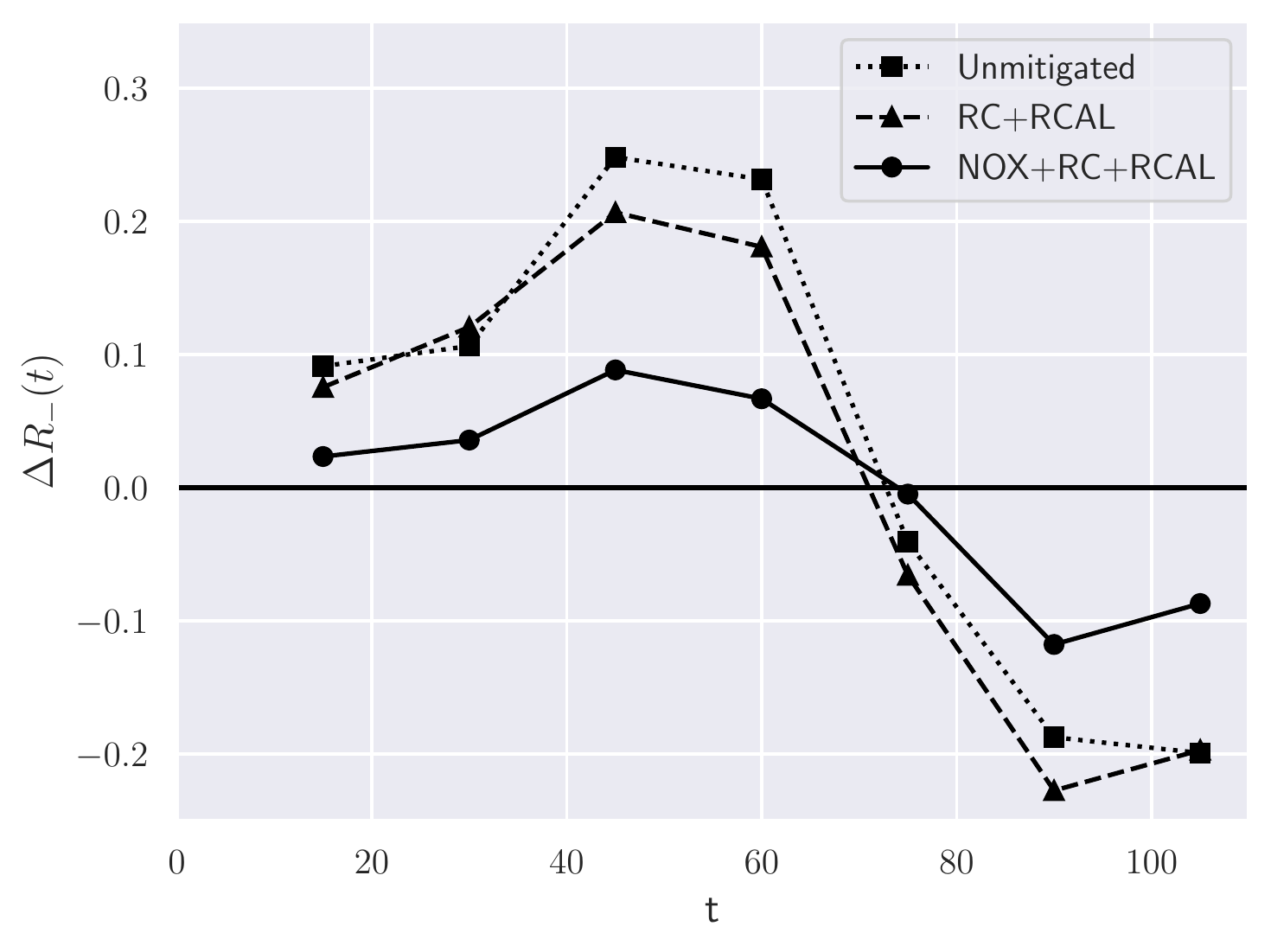}
     \caption{ free case }
     \label{fig:minus_free}
     \end{subfigure}
     \begin{subfigure}{8cm}
     \includegraphics[width=8cm]{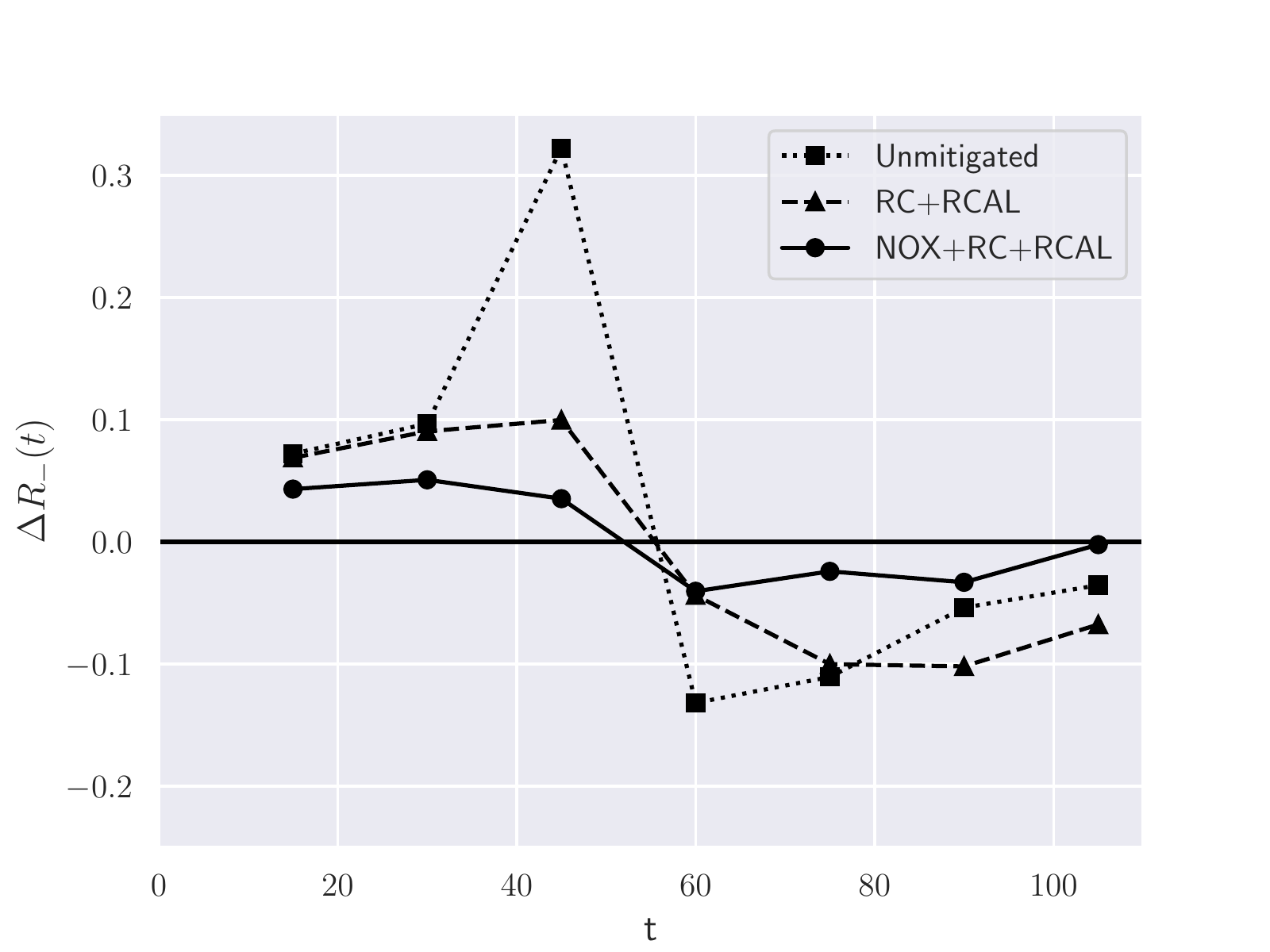}
    \caption{ interacting case }
     \label{fig:minus_int}
     \end{subfigure}
    \caption{Graph of difference between computed $R_{-}(t)$ and the ideal value versus trotter time step for unmitigated, RC+RCAL, and NOX+RC+RCAL \texttt{ibmq\_kolkata} free and interacting cases}
    \label{fig:minus}
 \end{figure}

The next analysis of the \texttt{ibmq\_kolkata} data plotted $\Delta R_{-}(t)$ versus t for the full Trotter evolution for each mitigation procedure in \cref{fig:minus}.  The graph clearly shows that for both the free (\cref{fig:minus_free}) and interacting (\cref{fig:minus_int}) cases, the NOX+RC+RCAL computations have smaller differences between the measured and ideal values than the RC+RCAL or the original unmitigated data. This set of graphs provides additional evidence that the NOX+RC+RCAL error mitigation significantly improves the computation of $R_{-}(t)$ versus t.

\begin{figure}[ht]
     \centering
     \begin{subfigure}{8cm}
     \includegraphics[width=8cm]{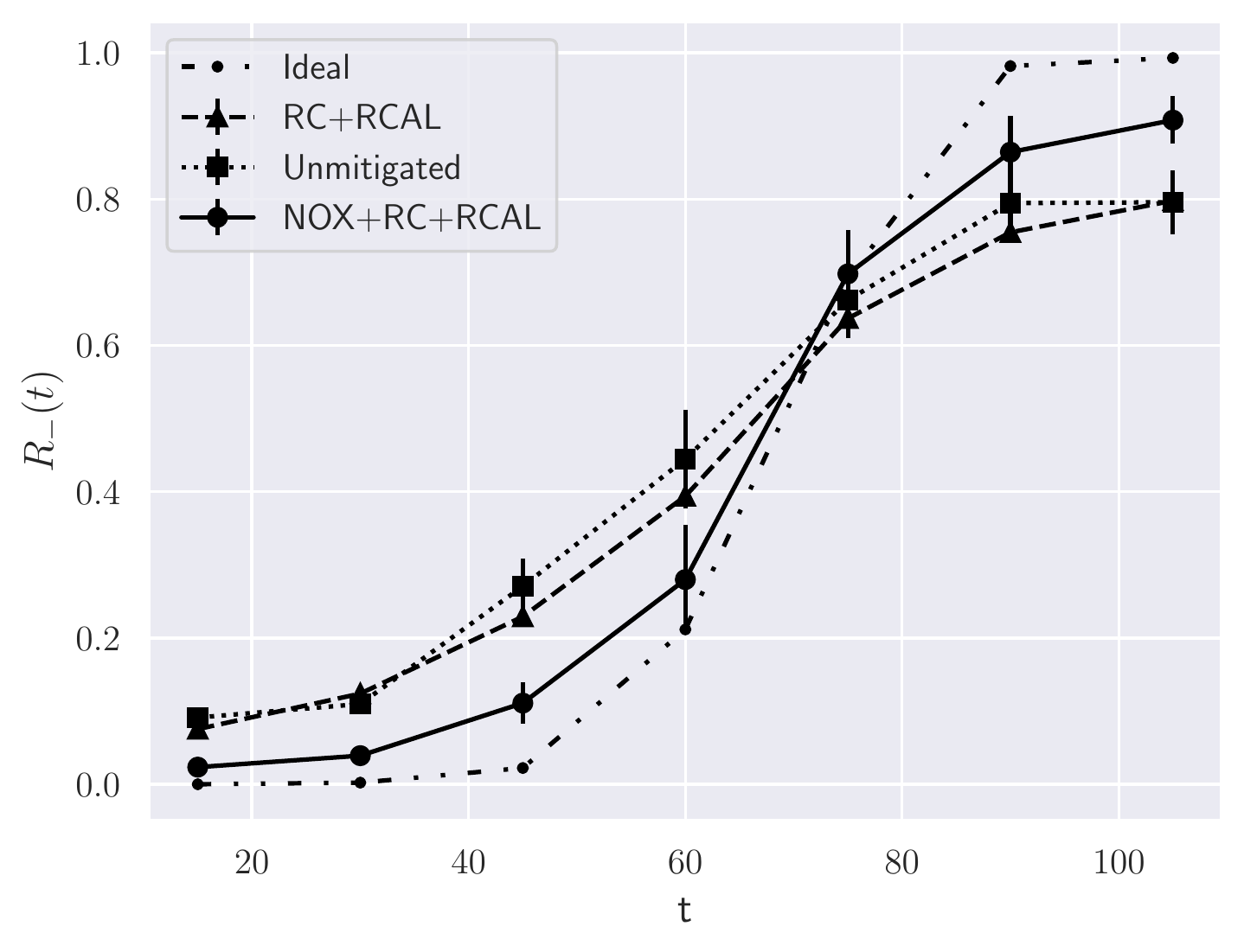}
     \caption{ free case }
     \label{fig:experiment_free}
     \end{subfigure}
     \begin{subfigure}{8cm}
     \includegraphics[width=8cm]{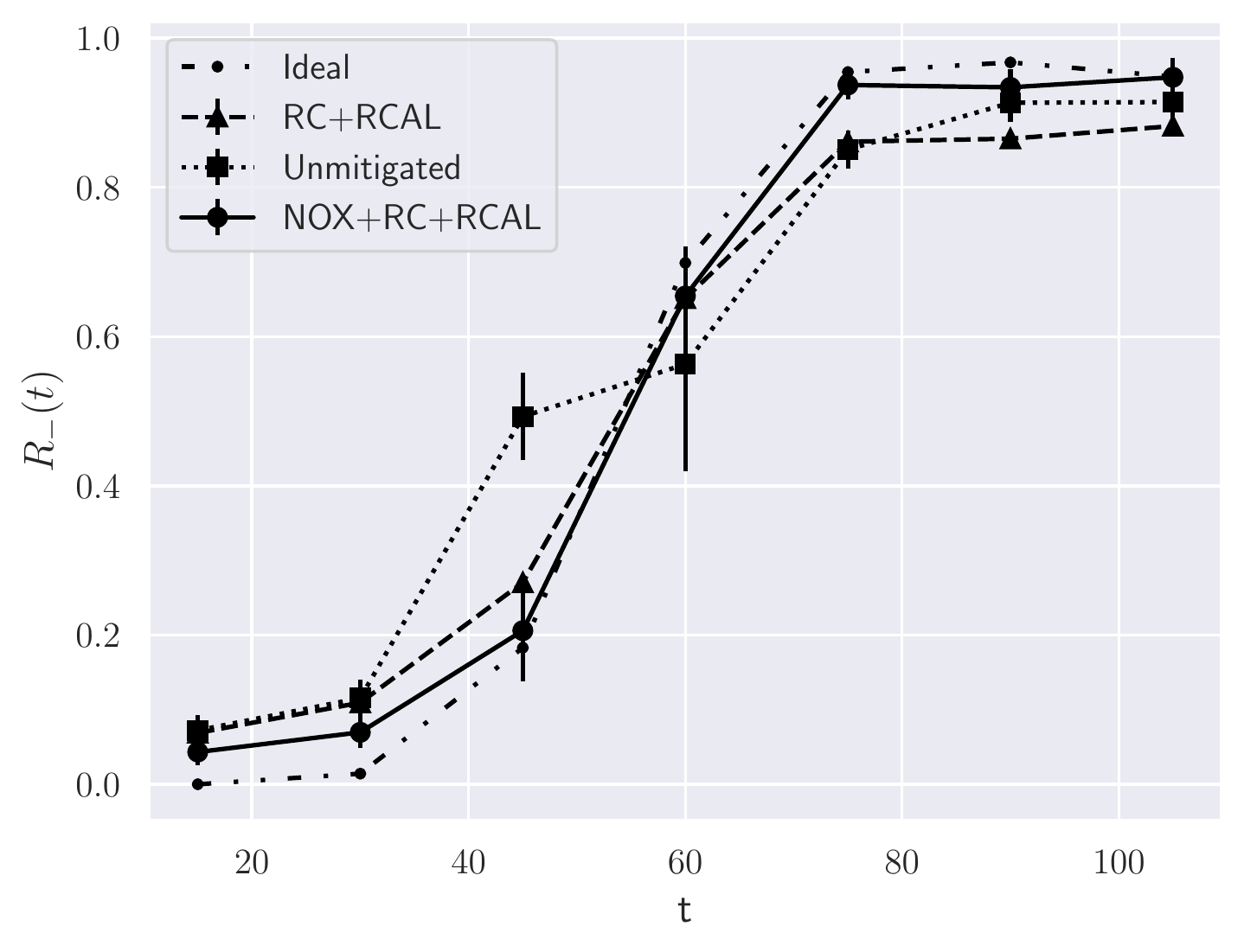}
    \caption{ interacting case }
     \label{fig:experiment_int}
     \end{subfigure}
    \caption{\texttt{ibmq\_kolkata} normalized reflection probabilities for the free and interacting cases for the ideal, unmitigated, RC+RCAL and NOX+RC+RCAL computations.  For both the free and interacting cases NOX combined with RC+RCAL showed substantial improvement to bring the \texttt{ibmq\_kolkata} $R_{-}t$ computations into close alignment with the ideally computed values.}
    \label{fig:experimental}
 \end{figure}

The \texttt{ibmq\_kolkata} data was next analyzed by plotting the normalized reflection probabilities versus $t$ for each mitigation procedure.  The results are plotted in \cref{fig:experimental}. For both the free and interacting cases, the plots of the RC+RCAL data showed some improvement compared to the unmitigated data.  The combination of NOX+RC+RCAL showed substantial improvement in the results above and beyond the RC+RCAL error mitigation.

\begin{figure}[ht]
     \centering
     \begin{subfigure}{8cm}
     \includegraphics[width=8cm]{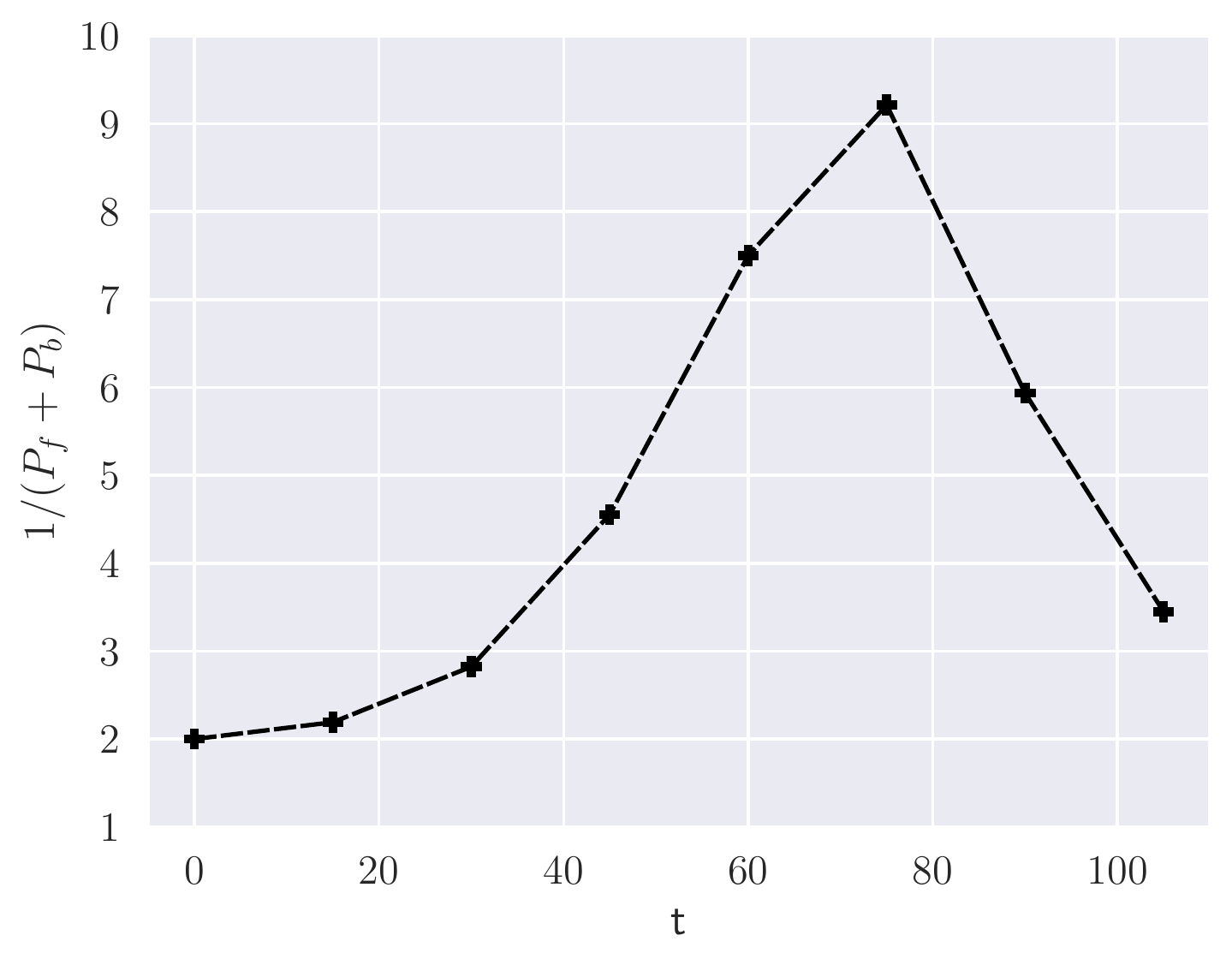}
     \caption{ Inverse probability, free case }
     \label{fig:inverseP_free}
     \end{subfigure}
     \begin{subfigure}{8cm}
     \includegraphics[width=8cm]{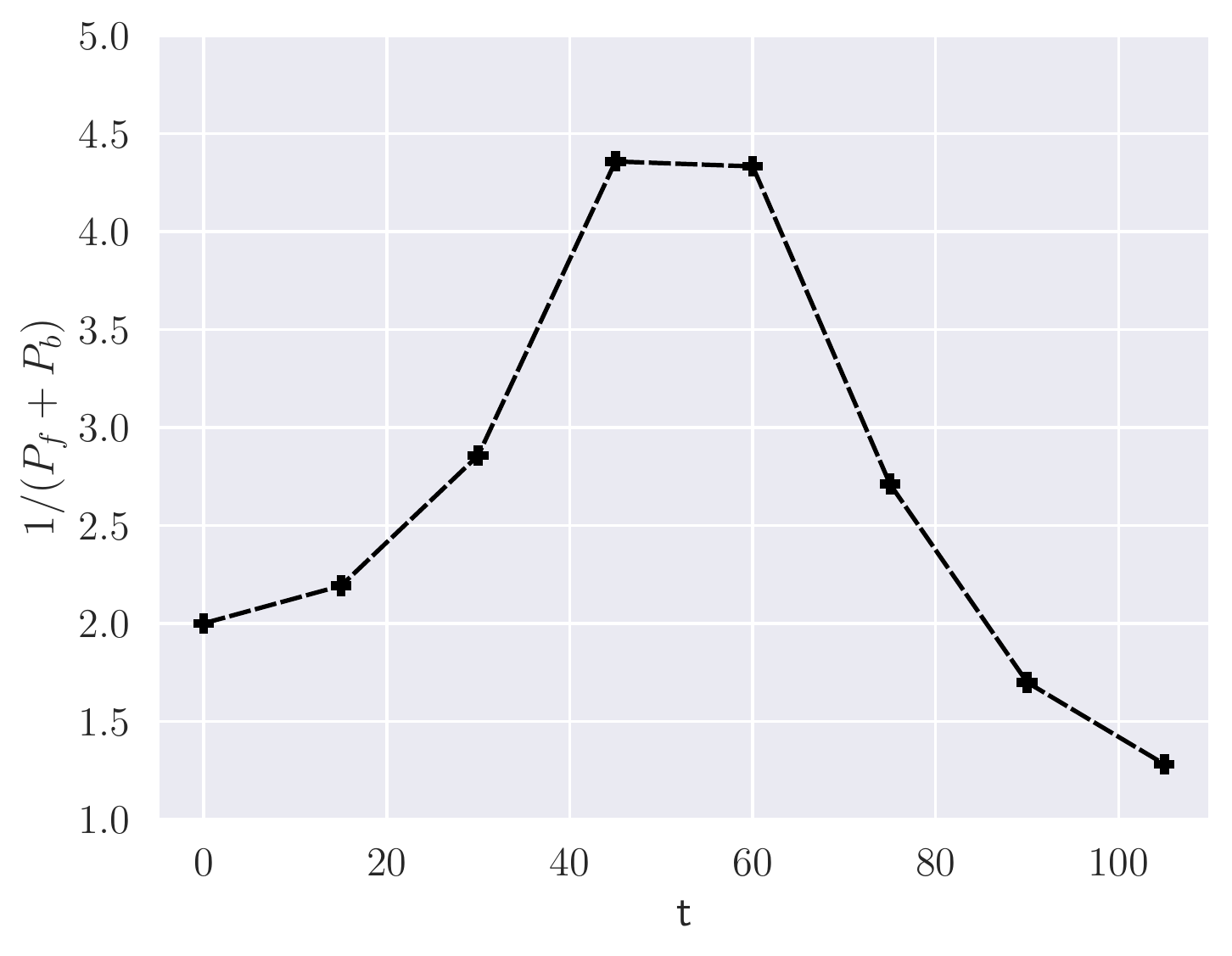}
    \caption{ Inverse probability, interacting case }
     \label{fig:inverseP_int}
     \end{subfigure}
    \caption{Ideal numerical calculations for $1/(P_{+}+P_{-})$ in the free and interacting cases.}
    \label{fig:inverseP}
 \end{figure}

We observed that the NOX values appear to be overestimated in the $\tilde{t}^*$ region where the $\Delta R_{-}(t)$ is changing rapidly.  We computed the ideal numerical calculations for $1/(P_++P_-)$  in the free and interacting cases as shown in  \cref{fig:inverseP_free,fig:inverseP_int}. We noticed that empirically that large values of $1/(P_++P_-)$ appear to be correlated with significant over estimations of the NOX errors. The peaks of $1/(P_++P_-)$, the change of slope in the graph and the largest error over estimations appear near the values of $\tilde{t}^*$ which have values of approximately 69 in the free case and 55 in the interacting case \cite{Gustafson_2021}.

\subsection{Metrics}
\label{sec:metrics}
To better quantify these observations, we introduce two metrics to define the closeness of the experimental results compared to the expected values:
    \begin{equation}
        M_{1}(\lbrace Q_{i,j} \rbrace) = \frac{1}{N}\sum_{i=1}^{N} \left(\frac{1}{7}\sum_{j=1}^7|Q_{i,j} - I_j|\right)
    \label{eq:M1}
    \end{equation}
    and
    \begin{equation}
        M_{2}(\lbrace Q_{i,j} \rbrace) = \sqrt{\frac{1}{7N}\sum_{i=1}^{N}\sum_{j=1}^7(Q_{i,j} - I_j)^{2}},
    \label{eq:M2}
    \end{equation}
    where $I_{j}$ represents the ideal Trotterization $R_{-}(t)$ values at Trotter step $j$ and $Q_{i,j}$ represents the measured $R_{-}(t)$ values at Trotter step $j$ at the $i^{th}$ experimental run. Referencing the post-processing analysis procedure in \cref{sec:Delta R}, note that $N=30$ total experimental runs for the unmitigated case and $N=1$ experimental runs for the NOX+RC+RCAL and RC+RCAL cases.
    
The values for these metrics obtained using the \texttt{ibmq\_kolkata} hardware platform are shown in \cref{tab:k-metrics}. The table shows that, for \texttt{ibmq\_kolkata}, the error reduction using M$_{1}$ is $61.05\%$ for the free case and $73.19\%$ for the interacting case. The $M_2$ metrics signal a similar improvement ($58.93\%$ for the free case  and  $75.28\%$ for the interacting case). 

The full analysis using these metrics was repeated for the data collected on the IBM quantum hardware platforms \texttt{ibmq\_guadalupe} and \texttt{ibmq\_manila}.  For the 16-qubit \texttt{ibmq\_guadalupe} device, we targeted qubits 7, 10, 12, 15, and on the 5-qubit \texttt{ibmq\_manila} device, we targeted qubits labeled 0, 1, 2, 3. The results from the analysis of the \texttt{ibmq\_guadalupe} data are shown in \cref{tab:g-metrics} and the \texttt{ibmq\_manila} data is shown in \cref{tab:m-metrics}.

We find that, across each device, the NOX data is consistently closer to the ideal values than the unmitigated data, although the total error reduction varies by device. The metrics obtained from the \texttt{ibmq\_kolkata} data show the greatest error reduction compared to the other two devices. This is expected because both \texttt{ibmq\_guadalupe} and \texttt{ibmq\_manila} were older less efficient processors compared to \texttt{ibmq\_kolkata}.  It is also noted that the interacting cases in each set of experiments show the largest improvement in the $M_1$ and $M_2$ metrics overall.

\renewcommand{\arraystretch}{1.3}
\begin{table*}
\centering
  \ttabbox[]{
    \begin{subfloatrow}
      \ttabbox{
        \begin{tabular}{|c|ccccc|c|}
            \hline\hline
            Metric & NOX+RC+RCAL & RC+RCAL & Unmit. & $\frac{\text{NOX+RC+RCAL}}{\text{Unmit.}}$ & $\frac{\text{RC+RCAL}}{\text{Unmit.}}$ & $(1-\frac{\text{NOX+RC+RCAL}}{\text{Unmit.}}) \times 100$\\\hline
            
            $M_1$ (free)        & $0.061(17)$ & $0.153(2)$ & $0.167(3)$ & 0.363(100) & 0.920(20) & 63.7 \\
            $M_2$ (free)        & $0.071(17)$ & $0.165(2)	$ & $0.187(3)$ & 0.382(93) & 0.883(18) & 61.8 \\
            $M_1$ (interacting) & $0.033(15)$ & $0.082(2)$ & $0.107(2)$ & 0.306(143) & 0.764(20) & 69.4 \\
            $M_2$ (interacting) & $0.036(16)$ & $0.084(2)$ & $0.135(2)$ & 0.267(117) & 0.627(16) & 73.3 \\\hline\hline
        \end{tabular}
        }{\caption{\texttt{ibmq\_kolkata}}
        \label{tab:k-metrics}}
        \vspace{0.2cm}
    \end{subfloatrow}
    \begin{subfloatrow}
      \ttabbox{
        \begin{tabular}{|c|ccccc|c|}
            \hline\hline
            Metric & NOX+RC+RCAL & RC+RCAL & Unmit. & $\frac{\text{NOX+RC+RCAL}}{\text{Unmit.}}$ & $\frac{\text{RC+RCAL}}{\text{Unmit.}}$ & $(1-\frac{\text{NOX+RC+RCAL}}{\text{Unmit.}}) \times 100$\\\hline
            
            $M_1$ (free)        & $0.102(15)$ & $0.172(3)$ & $0.194(2)$ & 0.527(78) & 0.885(17) & 47.3 \\
            $M_2$ (free)        & $0.115(18)$ & $0.181(3)$ & $0.211(3)$ & 0.543(86) & 0.858(16) & 45.7 \\
            $M_1$ (interacting) & $0.053(12)$ & $0.116(2)$ & $0.142(2)$ & 0.373(84) & 0.815(16) & 62.7 \\
            $M_2$ (interacting) & $0.058(13)$ & $0.120(2)$ & $0.156(2)$ & 0.372(82) & 0.768(15) & 62.8 \\\hline\hline
        \end{tabular}
        }{\caption{\texttt{ibmq\_guadalupe}}
        \label{tab:g-metrics}}%
        \vspace{0.2cm}
    \end{subfloatrow}
    \begin{subfloatrow}
      \ttabbox{
        \begin{tabular}{|c|ccccc|c|}
            \hline\hline
            Metric & NOX+RC+RCAL & RC+RCAL & Unmit. & $\frac{\text{NOX+RC+RCAL}}{\text{Unmit.}}$ & $\frac{\text{RC+RCAL}}{\text{Unmit.}}$ & $(1-\frac{\text{NOX+RC+RCAL}}{\text{Unmit.}}) \times 100$\\\hline
            
            $M_1$ (free)        & $0.205(6)$ & $0.258(2)$ & $0.275(2)$ & 0.745(24) & 0.939(12) & 25.5 \\
            $M_2$ (free)        & $0.230(7)$ & $0.268(3)$ & $0.295(3)$ & 0.779(25) & 0.909(12) & 22.1 \\
            $M_1$ (interacting) & $0.113(7)$ & $0.159(2)$ & $0.204(2)$ & 0.557(33) & 0.779(1) & 44.3 \\
            $M_2$ (interacting) & $0.124(6)$ & $0.164(2)$ & $0.218(2)$ & 0.571(30) & 0.751(10) & 43.9 \\\hline\hline
        \end{tabular}
      }{\caption{\texttt{ibmq\_manila}}
        \label{tab:m-metrics}}%
        \end{subfloatrow}
        
      }
      {\caption{Metrics $M_{1}$ and $M_{2}$ calculated for free and interacting cases on three, superconducting IBM Quantum devices: \texttt{ibmq\_kolkata}, \texttt{ibmq\_guadalupe}, and \texttt{ibmq\_manila}.}\label{tab:metrics}}
    \end{table*}

\subsection{Wigner $\Delta{t}$ phase shift calculation}
\label{sec:Wigner-time-shift}

The next analysis of the \texttt{ibmq\_kolkata} data focused on calculating the Wigner $\Delta{t}$ phase shift. The results from the reflection probability measurements using the NOX+RC+RCAL error mitigation protocol can be directly applied to the calculation of the phase shifts.  

It is noted that the NOX+RC+RCAL yields significant error reduction for scattering probabilities $P_{\pm}(t)$ to be in the $\dket{\pm k}$ state for the QIM \cite{Gustafson_2021}. As discussed in \cref{sec:physics-bkgnd}, the normalized reflection probability shown in \cref{eq:rpminus}  allows us to estimate time delays and phase shifts due to interactions.

\begin{figure}
    \centering
    \includegraphics[width=8cm]{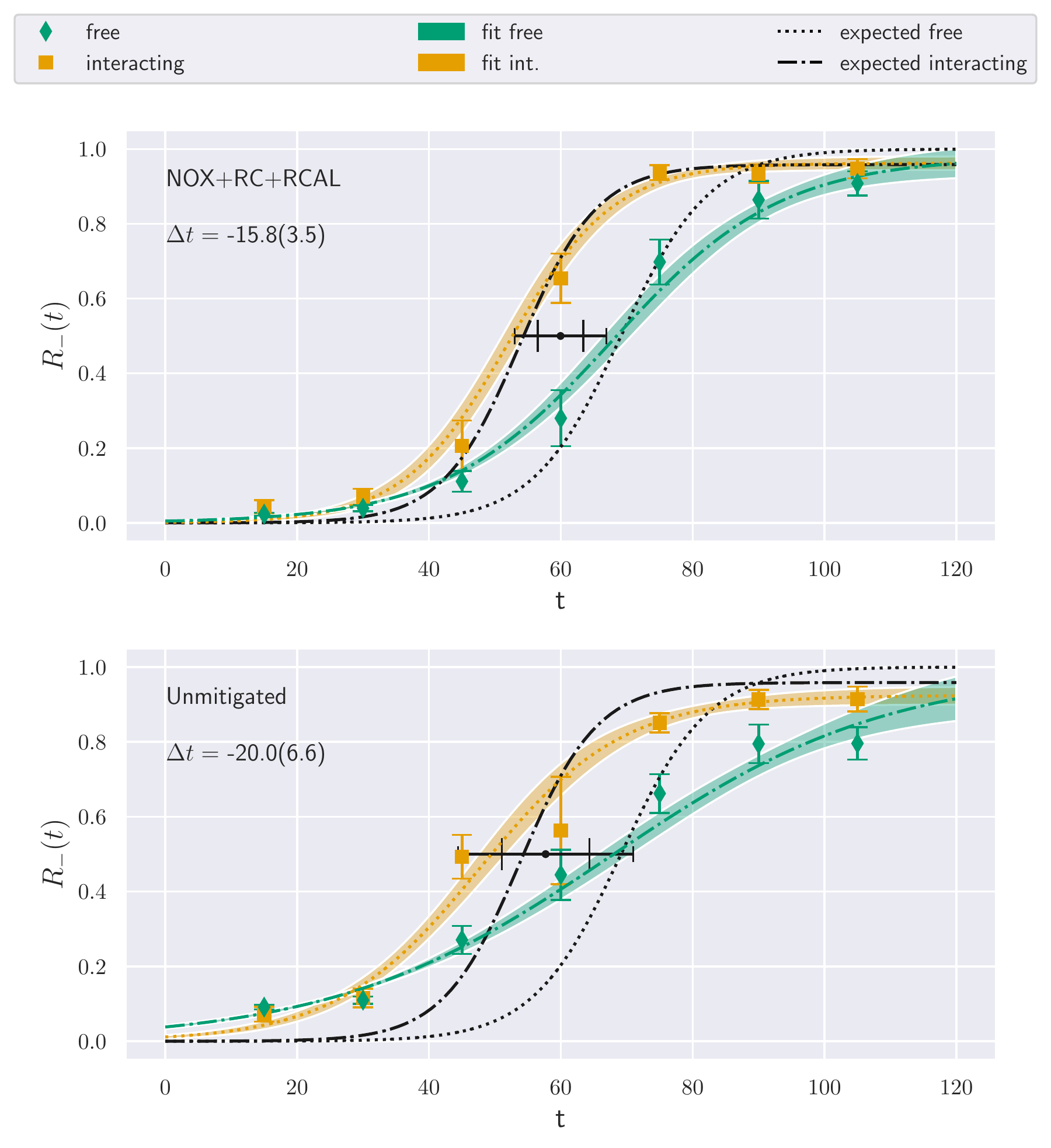}
    \caption{The time evolution and sigmoid fits for the simulation on \texttt{ibmq\_kolkata} including the NOX+RC+RCAL error mitigation and with only the unmitigated data.}
    \label{fig:examplenox}
\end{figure}

Following \cite{Gustafson_2021}, we assume the phase shift follows the empirical sigmoid function shown in \cref{eq:R-minus}. The sigmoid fits obtained using \cref{eq:R-minus} and the calculation of $\Delta t$ with and without NOX+RC+RCAL error suppression for the free case (\cref{fig:examplenox}a) and interacting case (\cref{fig:examplenox}b) were then computed.  The differences of $ \tilde{t}^*$ give us the time delay between the free and interacting wave packets. We define this difference as $\Delta t$.

By calculating and comparing the percentage differences between the experimental results ($\Delta t_{\rm Unmit.}$ and $\Delta t_{\rm NOX}$) and the ideal results, we can see that the accuracy of the $\Delta t_{\rm NOX}$  compared to $\Delta t_{\rm Unmit.}$ has improved by $28.71\%$ and that the error bar has decreased by $63.41\%$. While both results agree with the expected value of $\Delta t=-14.91(52)$, the NOX calculation shrinks the error bands and provides a more precise result.  The signs of the differences are similar for the two methods (positive for $t\leq 60$ and negative later). However, the magnitude of these difference are significantly smaller for the NOX procedure. Visually, it is clear that the NOX errors are about twice smaller than the unmitigated errors.

By design, the NOX mitigation technique produces more accurate expectation values for the observables under consideration.  These are the $|0100 \rangle \langle 0100|$ and $|1000 \rangle \langle 1000|$ states corresponding to $\pm k$ momentum states after the Fourier transform. Like other mitigation techniques, NOX trades precision for accuracy but compensates for the loss in precision by increasing the number of circuits and shots. Nonetheless, getting an error bar that adequately represents the precision of the mitigated expectation values remains a challenge because the assumptions underlying NOX (as well as other mitigation techniques) aren't guaranteed to apply perfectly, which may result in systematic loss of accuracy.

This is likely an artifact that the scattering process is elastic and far from any resonances. Together, these two factors render the scattering process semi-robust to noise, given that the particle does not slosh back and forth within the potential well. In this sense, the time delay $\Delta t$ is a good example of a physical quantity that is robust against computation errors and consequently weakly sensitive to error mitigation.  To estimate the time delay due to the interaction, it is very useful to have estimated systematic errors as discussed in \cref{sec:append_noxerror}.  This allows a determination of the free parameters using $\chi^2$ minimization.

To account for these effective systematic errors, we provide error bars on NOX-mitigated results by adding a systematic heuristic upper-bound to the statistical error bar obtained by taking the magnitude of the NOX correction~\cref{Eq:NOX-correct}:
\begin{align}
    \Delta^{\rm sys}_{\rm bound}:  = \left| \hat{O}_{\rm NOX+  RC} - \hat{O}_{\rm RC} \right|
\label{Eq:NOX-correct}
\end{align}
We point out that this upper bound is often generous for the systems under scope and purposefully overestimates systematic errors. 
Accurately tightening the systematic error bars of mitigated outcomes for computations vulnerable to non-Markovian or time-dependent noise sources remains an open problem.  We expand on the reasoning behind this heuristic in the Supplemental Material in \cref{sec:append_noxerror}.

\section{Summary}
\label{sec:summary}

We improved the previous real-time scattering calculations by applying three compounding error suppression techniques: RCAL, RC and NOX. These improvements are applicable and can be implemented across a spectrum of STEM application domains.  For a wide range of circuit depths and on three different devices, we consistently observed noticeable error reductions from NOX alone \emph{on top} of the error reduction provided by RC and RCAL.  However, the sensitivity to improved error mitigation results will likely be observable dependent.  This improved accuracy for our proof-of-concept application circuits demonstrated the applicability of NOX on cloud platforms, in the advent where circuits are executed in substantially time-separated batches. We further supplemented our mitigated results with systematic error bars that accounted for unmitigated errors. Future work is planned for further refinement of systematic errors.

We also note crucial differences between the NOX protocol and other noise extrapolation methods based on RIIM \cite{He2020ZNERIIM}. Although both RIIM and NOX implement similar approaches, the RIIM protocol targets individual cX gates.  Because RIIM is a noise-agnostic method, it cannot correctly amplify noise processes that do not commute with the cX gates.  NOX is a more noise-aware version of the standard RIIM noise extrapolation techniques. NOX targets entire gate cycles affected by various non-local and non-depolarising noise processes. NOX, therefore, provides a broader error suppression context that enables more accurate amplification of noise processes than methods that only focus on individual gates.

\begin{acknowledgments}

 Y. Meurice is supported in part by the U.S. Department of Energy (DoE) under Award Number DE-SC0019139. Erik Gustafson is supported by the DoE QuantISED program through the theory consortium ``Intersections of QIS and Theoretical Particle Physics" at Fermilab and by the U.S. Department of Energy. Fermilab is operated by Fermi Research Alliance, LLC under contract number DE-AC02-07CH11359 with the United States Department of Energy.  P. Dreher was supported in part by the U.S. Department of Energy (DoE) under award DE-AC05-00OR22725. This material is based upon work supported by the National Science Foundation Graduate Research Fellowship under Grant No. DGE-2137100. We thank North Carolina State University (NCSU) for access to the IBM Quantum Network quantum computing hardware platforms through the IBM Quantum Innovation Center at NC State University.  We acknowledge the use of IBM Quantum services for this work. The views expressed are those of the authors, and do not reflect the official policy or position of IBM or the IBM Quantum team.  Y.M and E.G. thank the members of QuLAT for suggestions and comments.  The project team acknowledges the use of True-Q software from Keysight Technologies \cite{trueq}.

\end{acknowledgments}

\section{Competing interests}
A. C-D. has a financial interest in Keysight Technology Inc. and the use of True-Q software \cite{trueq}. The remaining authors declare no competing interests.

 \section{Data Availability}
Data is available from the corresponding author upon reasonable request.


%

\vspace{5mm}

\appendix

\begin{center}

\Large
{\textbf{Supplementary Material}}
\end{center}
\normalsize

\section{Error Estimation Using Noiseless Output Extrapolation }
\label{sec:append_noxerror}

As shown in the previous section, NOX ideally induces a first-order correction on
the output. However, this mitigation effect is
founded around a Markovian time-independent
error model, as well as on a perfect error
amplification $\mc E_i \rightarrow \mc
E_i^{1+\alpha}$ through cycle repetition. A
number of physical mechanisms can bring us
outside of this framework, and reduce the accuracy of NOX-mitigated results.

For instance, one mechanism in which accuracy can be lost is if an error channel $\mc E_i$ doesn't exactly commute with the cycle $C_i$. In this case,
the amplified error resulting from repeating
$\E_i^{\rm amp} = C_i^{-(1+\alpha)}\left(\mc E_i C_i \right)^{1+\alpha}$,
might slightly differ from the amplified channel $\E_i^{1+\alpha}$. Other mechanisms for systematic loss in accuracy include time-dependent effects such as error drift and re-calibration, as well as non-Markovian effects.

To account for these effective systematic errors, we provide error bars on NOX-mitigated results by
adding a systematic heuristic upper-bound to the statistical error bar obtained by taking
the magnitude of the NOX correction:
\begin{align}
    \Delta^{\rm sys}_{\rm bound}:  = \left| \hat{O}_{\rm RC+  NOX} - \hat{O}_{\rm RC} \right|~.
\end{align}
To see the reasoning behind this choice of upper bound, recall \cref{{eq:nox_delta_horder}}.
To accommodate for faulty error amplification mechanisms, we modify \cref{{eq:nox_delta_horder}} by adding new terms:
\begin{align}\label{eq:nox_delta}
    \langle \mc C_i\rangle_{\rm RC}  = \mc C_{\rm ideal}  + \alpha \Delta_i +\alpha \Delta^{(1)}_i+\Delta^{(2)}_i+\sum_{j=1}^m \Delta_j + h.o.
\end{align}
Here the term $\alpha \Delta^{(1)}_i$ would represent the effect of a systematic fluctuation in the amplification mechanism; the $\Delta^{(2)}_i$ term would instead correspond to systematic fluctuations in non-amplified error components due to e.g time-dependent effects arising between circuits.  By substituting  ~\cref{eq:nox_delta} into ~\cref{eq:nox}
we obtain
\begin{align}
    \hat{O}_{\rm RC+  NOX} = \hat{O}_{\rm ideal} - \sum_i \mbb E \left ( O \Bigg| \Delta_i^{(1)}+\frac{\Delta^{(2)}_i}{\alpha} \right)  + h.o.
\end{align}
where the higher order terms are of second order in the circuit error probability.
Let's define the following error ratio
\begin{align}\label{eq:defgamma}
    \Gamma := \frac{\left| \sum_i \mbb E \left ( O | \Delta_i\right)\right|}{\left| \sum_i \mbb E \left ( O \Bigg| \Delta_i^{(1)}+\frac{\Delta^{(2)}_i}{\alpha}\right)\right|} 
\end{align}
The numerator is the first-order error term meant to be removed by NOX. The denominator is zero in the absence of violations to the framework that led to  \cref{{eq:nox_delta_horder}}.
With the expectation of weak violations (weaker than the ``baseline'' noise induced by $\Delta_i$ ) we certainly expect $\Gamma$ to be greater than 1.  Here, we consider the eventuality that the effect of leftover error $\sum_i\Delta_i^{(1)}+\frac{\Delta^{(2)}_i}{\alpha}$ on the observable $O$ is at least comparable to the effect of the first order unmitigated error $\sum_i\Delta_i$. That is, instead of assuming $\Gamma \gg 1$ (which holds when there are close to no violations of the assumptions that led to \cref{{eq:nox_delta_horder}}) we make the much more cautious assumption $\Gamma \geq 2$. In other words, we assume that effects such as imperfect averaging due to time-dependent error rate fluctuations aren't necessarily negligible, but are at least twice smaller than baseline error rates.
From there, we get our first-order upper bound on the
systematic error in NOX:
\begin{align}
    &\left| \hat{O}_{\rm RC+  NOX} - \hat{O}_{\rm RC} \right|  \simeq \left|\sum_i \mbb E \left ( O \Bigg| \Delta_i -\Delta_i^{(1)}-\frac{\Delta^{(2)}_i}{\alpha} \right) \right| \tag{\cref{eq:nox}, \cref{eq:rc_delta_appA}, \cref{eq:nox_delta}} \\
    & \geq \left|\left|\sum_i \mbb E ( O | \Delta_i ) \right| - \left| \sum_i  \mbb E \left( O\Big| \Delta_i^{(1)}+\frac{\Delta^{(2)}_i}{\alpha} \right) \right| \right| \tag{rev. triang. ineq.} \\
    & = |\Gamma-1| \left| \sum_i  \mbb E \left( O\Big| \Delta_i^{(1)}+\frac{\Delta^{(2)}_i}{\alpha} \right) \right| \tag{def. of $\Gamma$, \cref{eq:defgamma}} \\
    & \geq \left| \sum_i  \mbb E \left( O\Big| \Delta_i^{(1)}+\frac{\Delta^{(2)}_i}{\alpha} \right) \right|~.\tag{Assumption $\Gamma \geq 2$}
\end{align}

\begin{figure}[h]
     \centering
     \begin{subfigure}{8cm}
     \includegraphics[width=8cm]{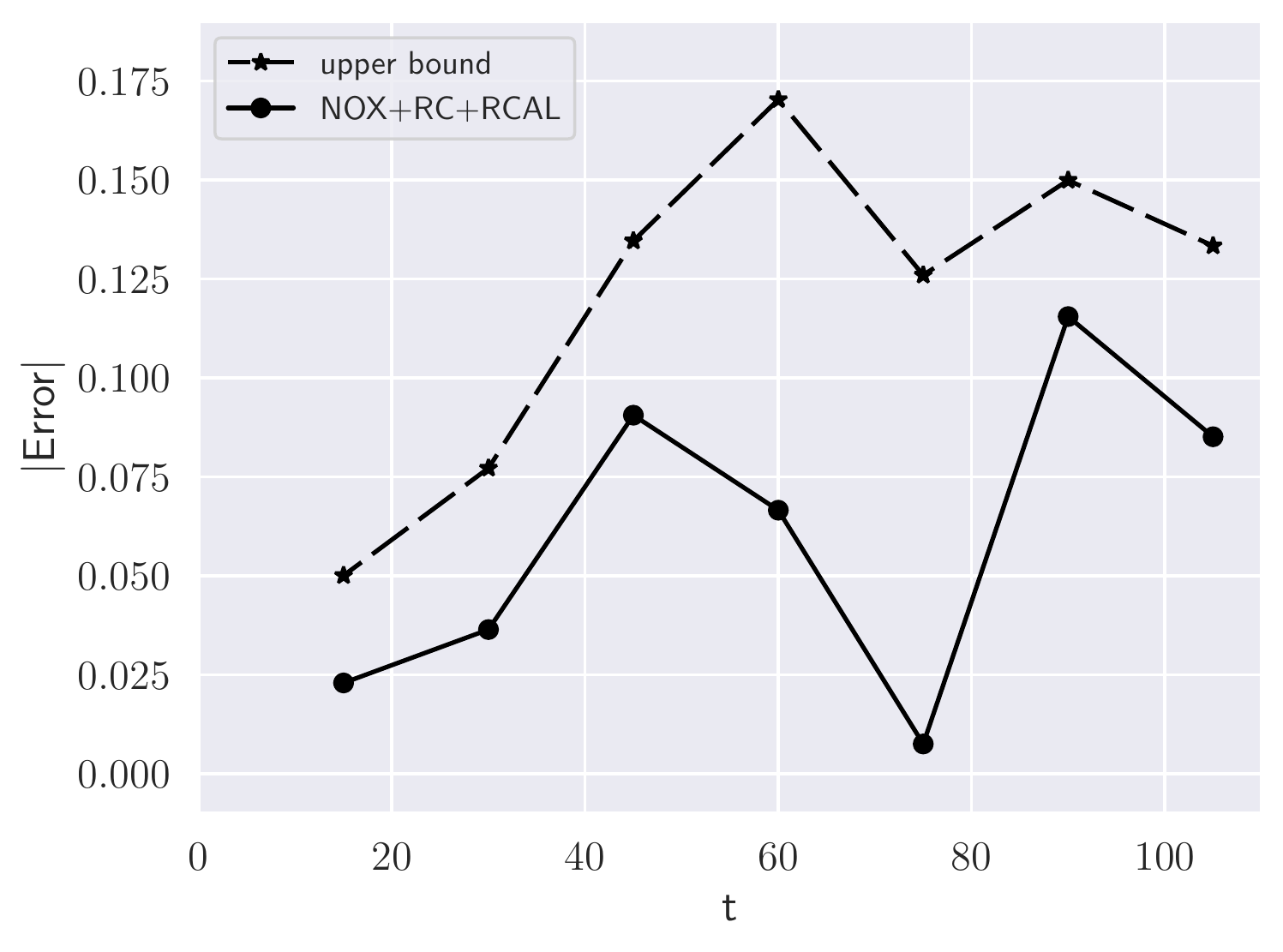}
     \caption{ free case }
     \label{fig:errorboundsfree}
     \end{subfigure}
     \begin{subfigure}{8cm}
     \includegraphics[width=8cm]{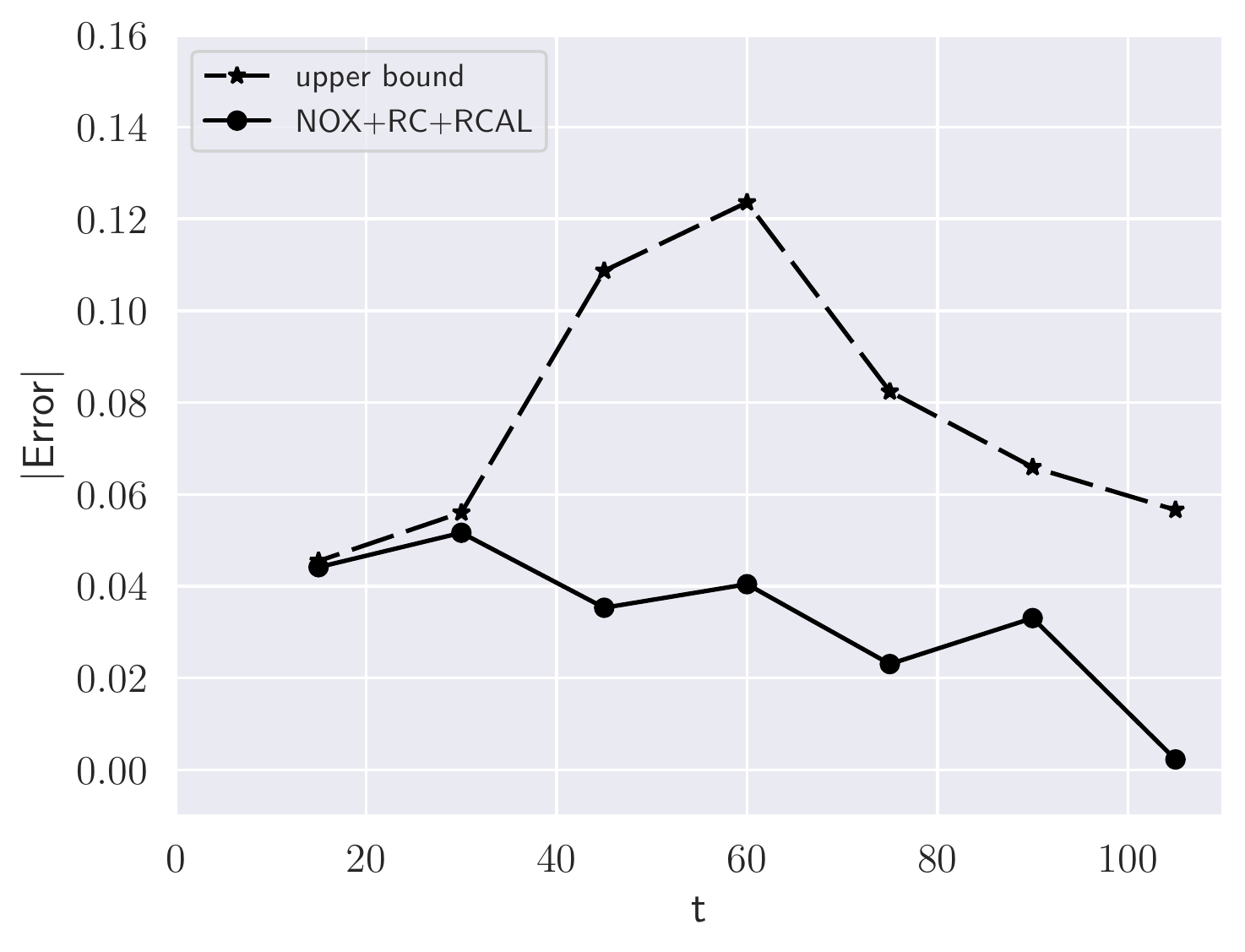}
    \caption{ interacting case }
     \label{fig:errorboundsint}
     \end{subfigure}
    \caption{Actual (solid line, circles) and heuristic upper bound (dashed line, stars) on NOX errors for calculated $R_-(t)$ values in absolute value for \texttt{ibmq\_kolkata} in the free and interacting cases. }
    \label{fig:estimatederrors}
 \end{figure}

{Note that the assumption $\Gamma \geq 2$ is coarse and inspired by empirical data.
(see \cref{fig:estimatederrors}).
Refining device-specific or application-specific upper bounds on $\Gamma$ is left as an open research avenue.}

\end{document}